\documentclass[
preprint2
]{aastex}

\usepackage[fleqn]{amsmath}
\usepackage{graphicx}
\usepackage[FIGBOTCAP]{subfigure}
\usepackage{txfonts}
\usepackage{expdlist}
\usepackage{multirow}
\usepackage{natbib}

\newcommand{\vONE}[1]{#1}

\bibpunct{(}{)}{;}{a}{}{,}

\title{On the simultaneous evolution\\of massive protostars and their host cores}

\author{R.~Kuiper}
\affil{
Jet Propulsion Laboratory,
California Institute of Technology,
4800 Oak Grove Drive,
Pasadena, CA 91109,
USA
}
\email{Rolf.Kuiper@jpl.nasa.gov}
\and
\author{H.~W.~Yorke}
\affil{
Jet Propulsion Laboratory,
California Institute of Technology,
4800 Oak Grove Drive,
Pasadena, CA 91109,
USA
}
\email{Harold.W.Yorke@jpl.nasa.gov}

\begin{abstract}
Studies of the evolution of massive protostars and the evolution of their host molecular cloud
cores are commonly treated as separate problems.  
However, interdependencies between the two can be significant.
Here, we study the simultaneous evolution of massive protostars and their host molecular cores using 
a multi-dimensional radiation hydrodynamics code that incorporates the effects of the thermal pressure
and radiative acceleration feedback of the centrally forming protostar.
The evolution of the massive protostar is computed simultaneously using the stellar evolution code STELLAR,
modified to include the effects of variable accretion.
The interdependencies are studied in three different collapse scenarios.
For comparison, stellar evolutionary tracks at constant accretion rates and the evolution of the host cores using pre-computed stellar evolutionary tracks are computed.

The resulting interdependencies of the protostellar evolution and the evolution of the environment are extremely diverse
and depend on the order of events, in particular the time of circumstellar accretion disk formation with respect to the onset of the bloating phase of the star.
Feedback mechanisms 
\vONE{affect}
the instantaneous accretion rate and the protostar's radius, temperature and luminosity on timescales $t \le 5 \mbox{ kyr}$, corresponding to the accretion timescale and Kelvin-Helmholtz contraction timescale, respectively.
Nevertheless, it is possible to approximate the overall protostellar evolution in many cases by pre-computed stellar evolutionary tracks assuming appropriate constant average accretion rates.
\end{abstract}

\keywords{
methods: numerical ---
stars: circumstellar matter ---
stars: evolution ---
stars: formation ---
stars: pre-main sequence
\\
\copyright 2012. All rights reserved
}

\begin{document}

\maketitle

\section{Introduction}
% Feedback of massive star onto their environment:
Massive luminous protostars strongly influence their environment by e.g.~thermal feedback, radiative acceleration feedback, protostellar winds, and ionization.
These various feedback mechanisms are of great interest in the context of clustered star formation and are investigated in numerical simulations \citep[e.g.][]{Yorke:2002p735, Krumholz:2007p416, Krumholz:2009p687, Peters:2010p16, Kuiper:2010p541, Peters:2011p394, Kuiper:2011p349, Cunningham:2011p953, Dale:2011p15667, Kuiper:2012p1151} 
and observations \citep[e.g.][]{Beuther:2002p3046, Kraus:2010p796, Wang:2011p3204, Wheelwright:2012p2430}.
Later in their evolution after dissipation of their accretion disks, massive stars will continue to influence their surroundings via e.g.~ionization, stellar winds and supernova explosions.
The interested reader is refered to \citet{MacLow:2004p350}, \citet{Beuther:2007p7046}, \citet{Zinnecker:2007p363}, and \citet{McKee:2007p456} for recent reviews on the field of (massive) star formation.

The physical state of a molecular core affects the accretion rate onto the embedded protostellar objects. 
The accretion rate onto the protostar itself is significantly 
%effected 
\vONE{affected}
by the formation of a circumstellar disk, by details of the angular momentum transport within the disk and by the launching of magnetic or radiation driven outflows, all of which in turn affect the surrounding molecular core.
Jets and outflows are ubiquitously observed in high-mass star forming regions \citep[e.g.][]{Shepherd:1996p577, Zhang:2001p711, Beuther:2002p3040, Beltran:2006p60, Zhang:2007p219, Fallscheer:2009p245, Leurini:2011p1160}.
Although direct observations of accretion disks in high-mass star forming regions are hampered by the fact that they are presumably short-lived, deeply embedded and at large distances, their existence can indirectly be inferred from the collimation of the jets and outflows \citep{Gomez:2003p3090, Brooks:2007p3194, Furuya:2008p673, Guzman:2012p4376}. 
Furthermore, the number of known disk candidates is rising \citep{Nielbock:2007p35, Bik:2008p509, Beuther:2008p734, Beuther:2009p580} and at least disk-like structures are detected around massive protostellar objects \citep{Bik:2004p636, Cesaroni:2005p272, Beuther:2005p655, Patel:2005p344, Chini:2006p288, Kraus:2010p796, Beuther:2010p359, Quanz:2010p4, Preibisch:2011p5363}.
The potentially best studied disk candidates are AFGL 490 \citep{Harvey:1979p392, Torrelles:1986p343, Chini:1991p339, Davis:1998p538, Lyder:1998p697, Schreyer:2002p769, Schreyer:2006p251} 
and IRAS 20126+4104 \citep{Wilking:1989p603, Cesaroni:1997p92, Zhang:1998p108, Cesaroni:1999p2140, Hofner:1999p2128, Cesaroni:2005p272, Sridharan:2005p1985}.
For the latest reviews on observational evidence for accretion disks around massive protostars, please see \citet{Zhang:2005p807} and \citet{Cesaroni:2007p8767}.

% `Simultaneous evolution', previous studies:
\citet{Stahler:1980p164} included the effect of accretion on the protostellar evolution for low-mass stars.
For intermediate-mass protostars, \citet{Palla:1991p4017, Palla:1992p3984} studied the evolution of protostars including the effect of different intermediate accretion rates.
\citet{Behrend:2001p774} compute the protostellar evolution of massive stars assuming growing (non-constant) accretion rates.
\citet{McKee:2003p282} present a simplified calculation of the evolution of the radius of a protostar and determine the protostellar accretion luminosity in a one-zone protostellar evolution model.
In Yorke \& Bodenheimer (2008, ASP Conf.Ser. 387, 189), \citet{Hosokawa:2009p23} and \citet{Hosokawa:2010p690} stellar evolutionary tracks for massive protostars are computed under the assumption of a time-independent constant accretion rate.
Using the pre-computed stellar evolutionary tracks by \citet{Hosokawa:2009p23} as a subgrid model for the forming massive protostar, \citet{Kuiper:2010p541} and \citet{Kuiper:2011p349} determined the evolution of their corresponding massive pre-stellar cores.
% Simultaneous evolution, this study:
Here we present and discuss the first simulations that take into account self-consistently the mutual feedback effects of an evolving massive protostar within the simultaneously evolving natal core.

\section{Methods}
\label{sect:methods}
We have incorporated the stellar evolution code STELLAR \citep{Bodenheimer:2007p501} modified by incorporating variable accretion \citep{Yorke:2008p1349} into our radiation-hydrodynamical framework for the evolution of the collapsing pre-stellar core \citep[introduced in][]{Kuiper:2010p586, Kuiper:2010p541, Kuiper:2011p349}.

\subsection{Solving for the evolution of the pre-stellar core}
To follow the evolution of the gas and dust of the collapsing pre-stellar core, we solve the equations of compressible radiation-hydrodynamics, including self-gravity and shear-viscosity, as described in \citet{Kuiper:2010p541}. 
For the basic hydrodynamics solver, we use the open source MHD code Pluto, described in \citet{Mignone:2007p544}.
As a special feature, our radiation transport solver takes into account the radiation pressure feedback by the forming protostar via a hybrid radiation transport approach, i.e.~using a frequency-dependent ray-tracing step for the stellar irradiation and a gray flux-limited diffusion approximation for the thermal dust emission.
Recently, in \citet{Kuiper:2012p1151}, we depicted the importance of the direct stellar irradiation feedback during the formation of massive stars regarding the stability of radiation-pressure-dominated cavities forming in the bipolar direction.
The derivation, the numerics, and tests of this hybrid scheme are given in \citet{Kuiper:2010p586}.
\vONE{
In \citet{Kuiper:2013p19029}, the reliability of this hybrid approach, the gray approximation, and the flux-limited diffusion approximation is compared to Monte-Carlo radiation transport results for circumstellar disks of a wide range of optical depths.
}

Here, we restrict ourselves to studies of the evolution of pre-stellar cores in axial symmetry only, which allows us to vary important parameters over a wide range.
As we shall show, the main impact on the protostellar accretion rate will be given by the formation of an accretion disk and the launching of bipolar outflows. 
Hence, the restriction to axial symmetry does not denote any limitation to the feedback mechanisms onto the actual protostellar evolution.
In contrast to our previous studies, the evolution of the protostar and the evolution of the pre-stellar core are computed simultaneously.

\subsection{Solving for the evolution of the protostar}
The textbook \citet{Bodenheimer:2007p501} includes a CD-ROM with the program package STELLAR, which implicitly solves the equations of stellar structure in spherical symmetry via the Henyey method, assuming a gray atmosphere. 
Convection is treated in the context of mixing-length theory. 
A nuclear burning reaction network for pp- and CNO-hydrogen burning and helium burning up to the 3-$\alpha$ process is included.  
The atmosphere module of STELLAR has been modified to allow ``cold'' accretion of material \citep{Yorke:2008p1349} onto the atmosphere, i.e.~it is assumed that most of the heat produced in an accretion shock is radiated away, before it settles onto the star.  
The accretion luminosity $L_{\mathrm acc}$ is accounted for separately, i.e.~$L_{\mathrm tot} = L_* + L_{\mathrm acc}$, whereby
\begin{equation}
L_\mathrm{acc} = \eta \frac{G~M_*}{R_*}~\dot{M}_* \; ,
\label{eq-acc}
\end{equation}
where $L_*$, $M_*$, $R_*$, and $\dot M_*$ are the (proto)stellar luminosity, mass and radius and the
mass accretion rate onto the (proto)star, respectively. 
$\eta$ is a dimensionless parameter of order unity but less than one, which takes into account the fact that not all of the accretion's kinetic energy is converted into radiative energy.  
For simplicity we use $\eta=1$.

As a starting model, we use a $0.05 \mbox{ M}_\odot$ object in hydrostatic equilibrium with a radius of $0.56 \mbox{ R}_\odot$ and a luminosity of $2.2 \times 10^{-2} \mbox{ L}_\odot$.

\section{Simulations overview, initial conditions, and numerical configuration}
\begin{table*}[tbhp]
\begin{tabular}{l | c c c c c | c c c c}
Tag & Stellar & Core & $\beta$ & $E_\mathrm{rot}/E_\mathrm{grav}$ & $\dot{M}_*$ & $R_*^\mathrm{max}$ & $M_*^\mathrm{Disk}$ & 
$M_*^\mathrm{Bloating}$ & 
$M_*^\mathrm{ZAMS}$ \\
& evolution & evolution & $\rho \propto r^{\beta}$ & [\%] & $\mbox{ [} 10^{-4} \mbox{ M}_\odot \mbox{ yr}^{-1} \mbox{]}$ & $\mbox{ [R}_\odot \mbox{]}$ & $\mbox{ [M}_\odot \mbox{]}$ & $\mbox{ [M}_\odot \mbox{]}$ &
$\mbox{ [M}_\odot \mbox{]}$ \\
\hline
RHD+SE A& Yes 	& Yes	& $-2.0$	& 1.0	& - & 885	& 22 	& 
5.0 - 22.0 & 
40 \\
RHD+SE B& Yes 	& Yes	& $-1.5$	& 1.7	& - & 456	& 7		& 
3.9 - 12.0 & 
30 \\
RHD+SE C& Yes 	& Yes	& $-1.5$	& 5.0	& - & 389 & 2.4 	&
3.5 - 10.7 & 
21 \\
\hline
RHD 	A	& No 	& Yes	& $-2.0$	& 1.0	& - & - 	& 25 	& - & - \\
RHD 	B	& No 	& Yes	& $-1.5$	& 1.7	& - & - 	& 7	 	& - & - \\
RHD 	C	& No 	& Yes	& $-1.5$	& 5.0	& - & - 	& 3.7	& - & - \\
\hline
SE 		& Yes	& No		& - 		& - 		& 30	& 1366 & - & 5.3 - 39.2 & 70 \\
SE A		& Yes	& No		& -		& - 		& 20 & \hspace{1ex}845 & - & 4.6 - 23.0 & 52 \\
SE 		& Yes	& No		& -		& - 		& 10 & \hspace{1ex}498 & - & 3.7 - 13.0 & 32 \\
SE B		& Yes	& No		& - 		& - 		& \hspace{1ex}9	& \hspace{1ex}462 & - & 3.6 - 12.2 & 30 \\
SE C		& Yes	& No		& -		& - 		& \hspace{1ex}7	& \hspace{1ex}377 & - & 3.5 - \hspace{1ex}9.9 & 25 \\
SE		& Yes	& No		& -		& - 		& \hspace{1ex}5	& \hspace{1ex}286 & - & 3.3 - \hspace{1ex}7.7 & 21 \\
SE		& Yes	& No		& -		& - 		& \hspace{1ex}1	& \hspace{2ex}31 & - & 2.6 - \hspace{1ex}3.9 & 10
\end{tabular}
\caption{
Overview of simulations performed.
From left to right the columns denote 
the run tag, 
the code configuration, and 
whether or not the protostellar evolution and the evolution of the pre-stellar core are computed.  
For cases when the evolution of the molecular core is computed, the next two columns denote
the initial density slope $\beta$ of the pre-stellar core and
the initial ratio of rotational to gravitational energy $E_\mathrm{rot}/E_\mathrm{grav}$ of the pre-stellar core.
The assumed constant accretion rate $\dot{M}_*$ given in the next column is only relevant when the
evolution of the molecular core is not computed.  
The last four columns denote
the resulting maximum protostellar radius $R_*^\mathrm{max}$,
the resulting stellar mass $M_*^\mathrm{Disk}$, at which the (Keplerian) circumstellar disk is formed,
the resulting stellar mass range $M_*^\mathrm{Bloating}$, 
%vONE in which the protostar remains bloated
\vONE{in which the protostar bloats up to its maximum radius,}
and the resulting stellar mass $M_*^\mathrm{ZAMS}$, at which the protostar reaches the zero-age-main-sequence.  
Dashes are given when these parameters are not relevant for a particular sequence.
}
\label{tab:runs}
\end{table*}
We performed three different types of numerical simulation:
The main study consists of simulations taking into account the simultaneous evolution of the protostar within its collapsing pre-stellar host molecular core.
These runs are labeled with ``RHD+SE'' (Radiation-Hydrodynamics + Stellar Evolution).
We consider three different initial conditions of the pre-stellar core morphology, labeled ``A'', ``B'', and ``C'', for which we vary the initial density slope and the ratio of rotational to gravitational energy.
For comparison purposes we perform a second type of simulation, calculating the evolution of pre-stellar cores -- with the three different initial conditions -- using the pre-computed stellar evolutionary tracks by \citet{Hosokawa:2009p23}.
\vONE{
These evolutionary tracks of massive stars depend on the stellar mass as well as on the actual accretion rate. 
We use polynomial fits to the mass relation up to tenth order for separated mass ranges and linear regression for the dependence on the instantaneous accretion rate, see \citet{Kuiper:2010p541} for details.
}
The numerical framework of these runs is consistent with our previous simulations \citep{Kuiper:2010p541, Kuiper:2011p349, Kuiper:2012p1151} and the runs are labeled with ``RHD''.
The third type of simulation, also for the purpose of comparison, is computing stellar evolutionary tracks at given constant accretion rates, labeled ``SE''.

The initial condition of the protostellar evolution is given by a $0.05 \mbox{ M}_\odot$ object in hydrostatic equilibrium with a radius of $0.56 \mbox{ R}_\odot$ and a luminosity of $2.2 \times 10^{-2} \mbox{ L}_\odot$.
The initial condition of the pre-stellar cores is given by 
an outer core radius of $0.1~\mbox{pc}$, 
a core mass of $100 \mbox{ M}_\odot$, and 
a core temperature of $20~\mbox{K}$.
The core is initially in rigid solid body rotation.
Our computational grid employs spherical coordinates $(r,\theta)$ assuming axial symmetry with a non-constant radial distribution of grid points and a central sink cell of $10~\mbox{AU}$.
The resolution of the grid at the inner computational boundary is about $1~\mbox{AU}$ in the radial as well as in the polar direction.
The polar extent of the grid covers the angles from $0\degr$ to $90\degr$, assuming equatorial symmetry.
Towards the outer computational boundary, the grid cell size increases linearly with radius.
For more details on the numerical grid in use, the numerical solvers, and the sub-grid models such as shear viscosity and the dust model, please see \citet{Kuiper:2010p586} and \citet{Kuiper:2010p541}.

An overview of all simulations performed is given in Table~\ref{tab:runs}.

The outstanding event during the protostellar evolution -- for non-negligible accretion -- is its bloating phase.
The increase and decrease of the stellar radius before and after the bloating phase, respectively, have a strong impact on the protostellar environment.
The (first) outstanding event during the pre-stellar core collapse is the formation of a circumstellar accretion disk.
To investigate the possible interdependencies of the protostellar and the host core evolution, we computed three different collapse scenarios, in which the circumstellar disk forms either at the end (run ``A''), during (run ``B''), or before (run ``C'') the protostellar bloating phase.

\section{Results}
In this section we discuss the main outcome of the protostellar and the pre-stellar core evolution in the three different collapse scenarios in light of four main evolutionary phases: 
(I) pre-bloating, 
(II) bloating, 
(III) Kelvin-Helmholtz contraction, and 
(IV) main sequence, which we shall discuss in more detail below. 
Note that \citet{Hosokawa:2009p23} and \citet{Hosokawa:2010p690} use the terms ``convection'' and ``swelling'' for phases (I) and (II).  
The four phases of evolution presented here are consistent with the ``cold disk accretion'' evolution by 
\citet{Hosokawa:2010p690}, but in contrast to their assumed constant accretion rates, we use the accretion rate computed self-consistently from the evolution of the collapsing pre-stellar core, including the radiation pressure feedback of the forming star.

\begin{figure}[htbp]
\begin{center}
\includegraphics[width=0.45\textwidth]{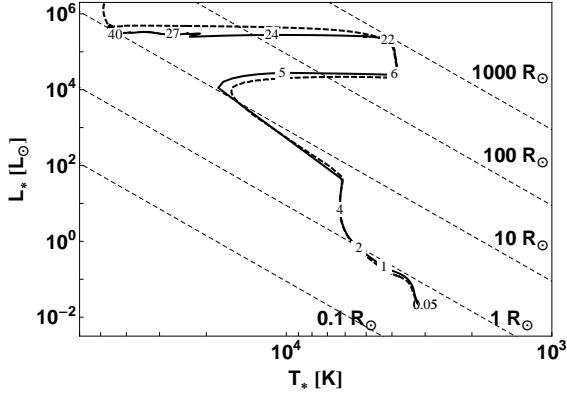}\\
\caption{
Hertzsprung-Russell diagram for the protostellar evolution in the collapse scenario ``RHD+SE A'' (solid line).
The dashed stellar evolutionary track denotes the evolution of the protostar in a comparison run assuming a constant accretion rate of $2\times10^{-3} \mbox{ M}_\odot \mbox{ yr}^{-1}$.
The diagonal dashed lines denote lines of constant stellar radii as labeled.
The small numbers along the stellar evolutionary track mark the protostellar mass.
}
\label{fig:A-HRD}
\end{center}
\end{figure}

\begin{figure}[htbp]
\begin{center}
\includegraphics[width=0.45\textwidth]{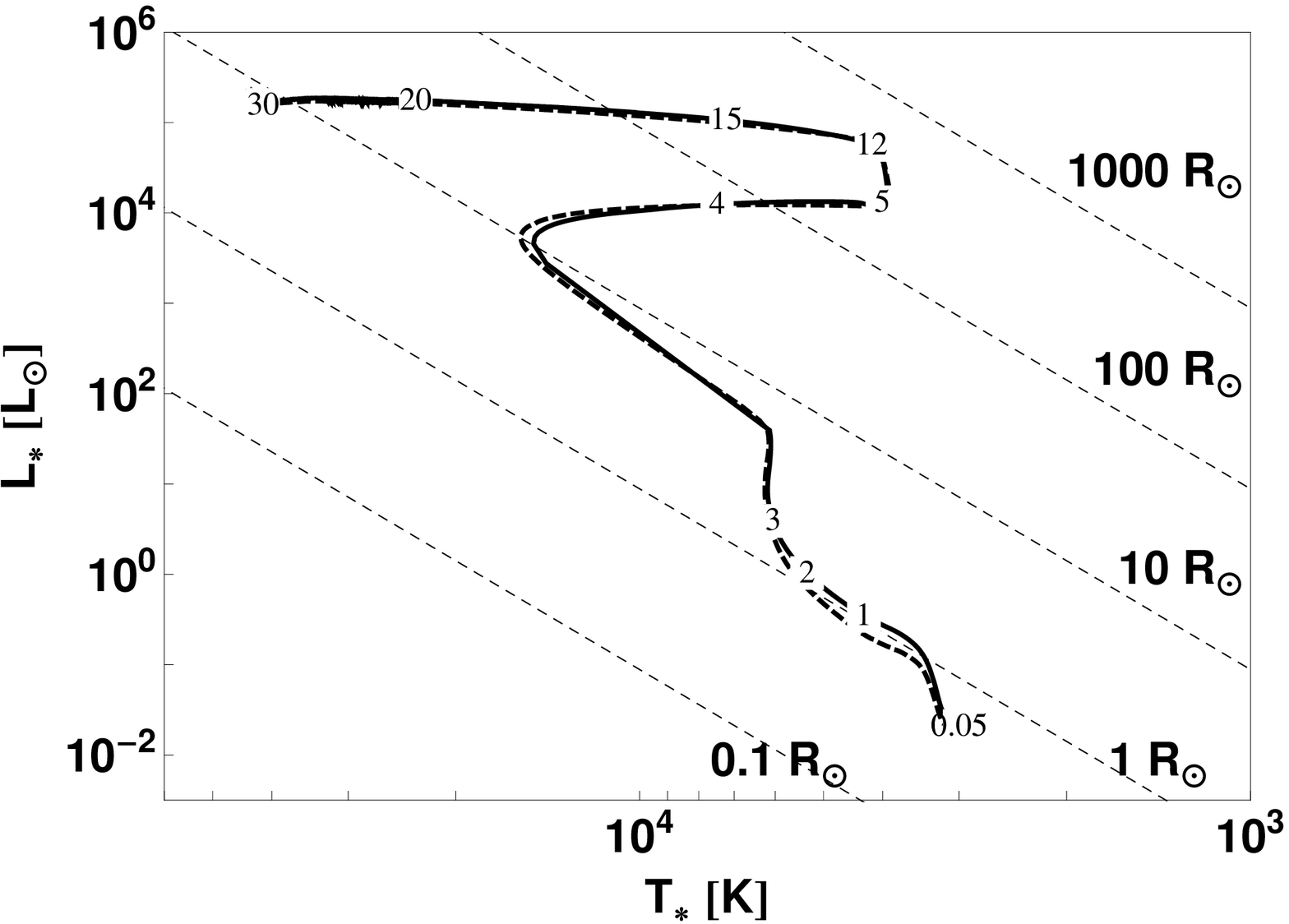}\\
\caption{
Hertzsprung-Russell diagram for the protostellar evolution in the collapse scenario ``RHD+SE B''.
The dashed stellar evolutionary track denotes the evolution of the protostar in a comparison run assuming a constant accretion rate of $9\times10^{-4} \mbox{ M}_\odot \mbox{ yr}^{-1}$.
Otherwise, symbols and numbers are as in Fig.~\ref{fig:A-HRD}. 
}
\label{fig:B-HRD}
\end{center}
\end{figure}

\begin{figure}[htbp]
\begin{center}
\includegraphics[width=0.45\textwidth]{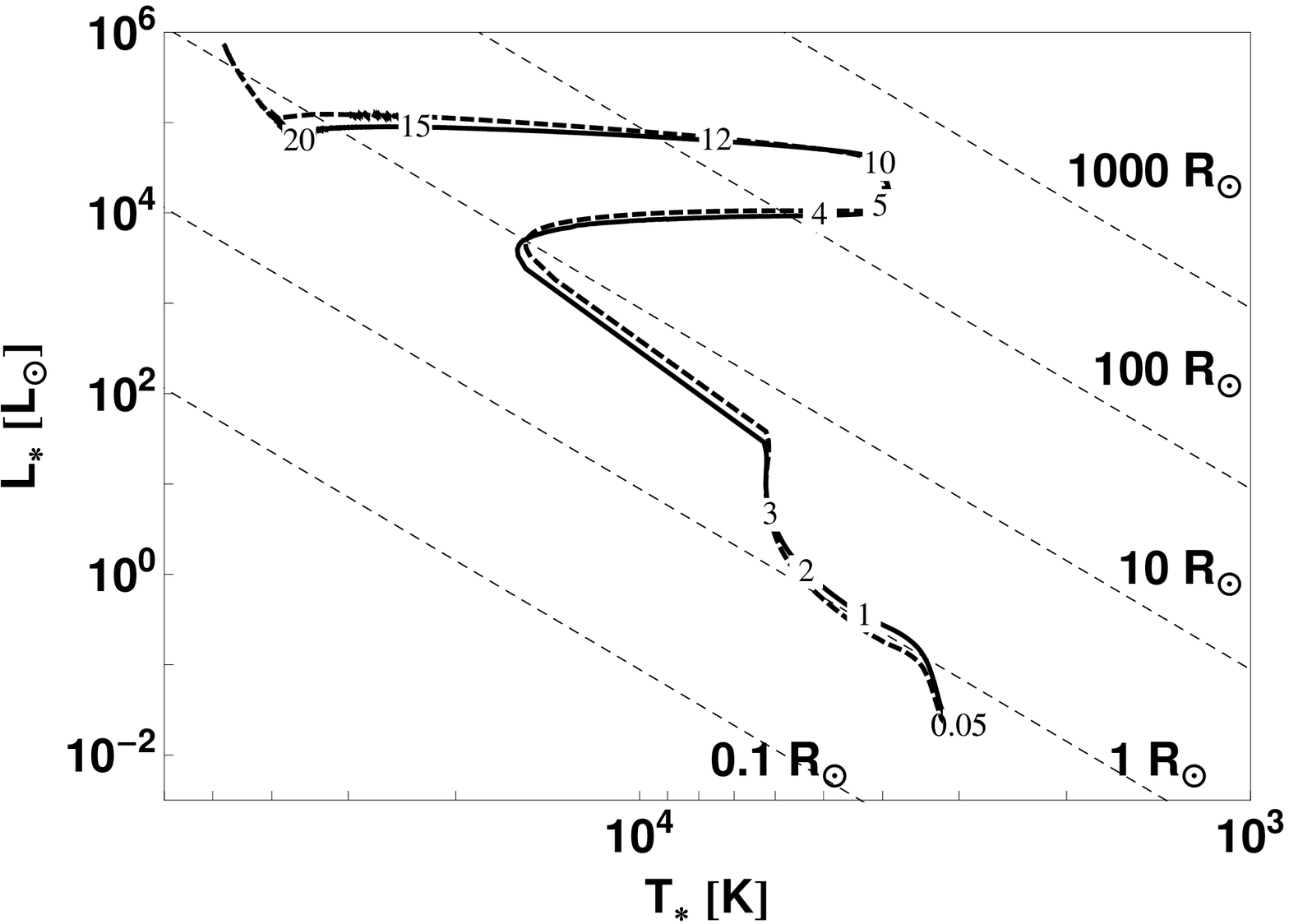}\\
\caption{
Hertzsprung-Russell diagram for the protostellar evolution in the collapse scenario ``RHD+SE C''.
The dashed stellar evolutionary track denotes the evolution of the protostar in a comparison run assuming a constant accretion rate of $7\times10^{-4} \mbox{ M}_\odot \mbox{ yr}^{-1}$.
Otherwise, symbols and numbers are as in Fig.~\ref{fig:A-HRD}. 
}
\label{fig:C-HRD}
\end{center}
\end{figure}

\begin{figure}[htbp]
\begin{center}
\includegraphics[width=0.45\textwidth]{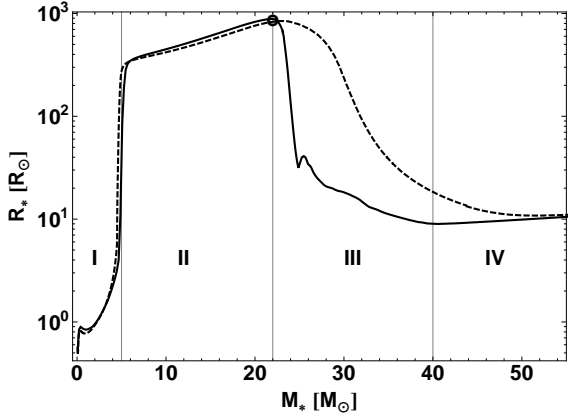}\\
\caption{
Evolution of the protostellar radius as a function of the protostellar mass in the collapse scenario ``RHD+SE A'' (solid line).
The circle (at 22$\mbox{ M}_\odot$) denotes the onset of the formation of the circumstellar disk in the computational domain.
The four phases: (I) pre-bloating, (II) bloating, (III) Kelvin-Helmholtz contraction, and (IV) main sequence evolution are as indicated.
The dashed line denotes the evolution of the comparison run ``SE A'' assuming a constant accretion rate of $2\times10^{-3}  \mbox{ M}_\odot \mbox{ yr}^{-1}$.
}
\label{fig:A-Rstar_vs_Mstar}
\end{center}
\end{figure}

\begin{figure}[htbp]
\begin{center}
\includegraphics[width=0.45\textwidth]{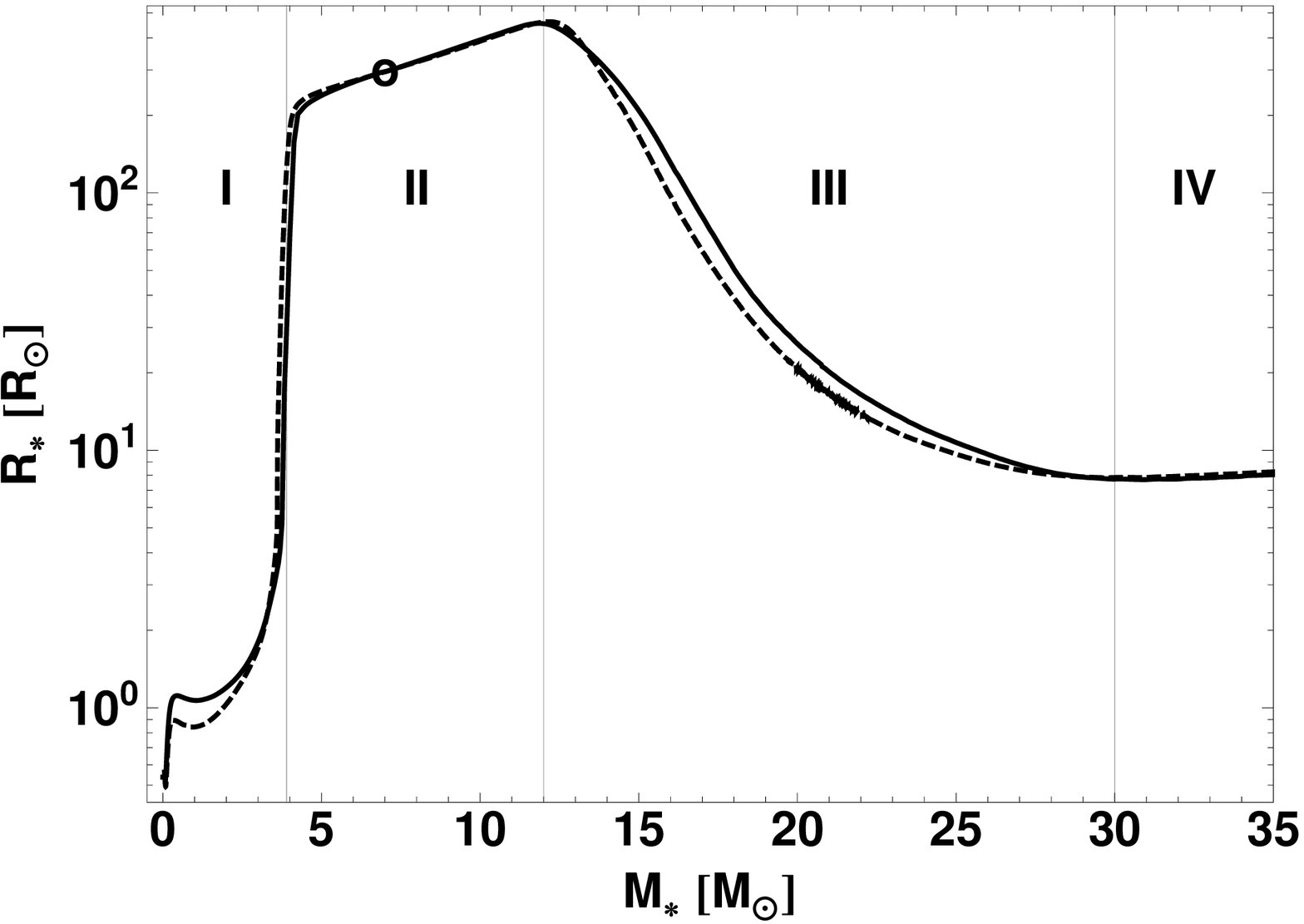}\\
\caption{
Evolution of the protostellar radius as a function of the protostellar mass in the collapse scenario ``RHD+SE B''; the four phases of evolution are as indicated.
The circle (at 7$\mbox{ M}_\odot$) denotes the onset of the formation of the circumstellar disk in the computational domain.
The dashed line denotes the evolution of the comparison run ``SE B'' assuming a constant accretion rate of $9\times10^{-4}  \mbox{ M}_\odot \mbox{ yr}^{-1}$.
}
\label{fig:B-Rstar_vs_Mstar}
\end{center}
\end{figure}

\begin{figure}[htbp]
\begin{center}
\includegraphics[width=0.45\textwidth]{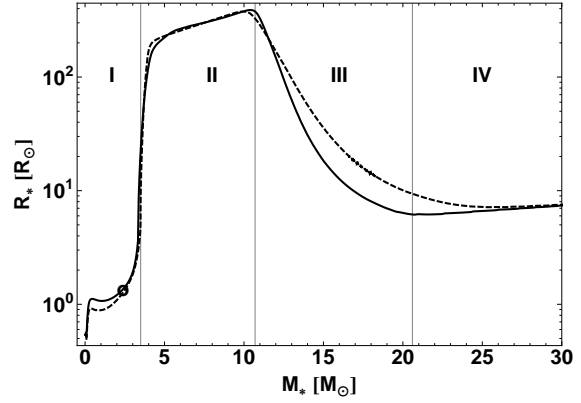}\\
\caption{
Evolution of the protostellar radius as a function of the protostellar mass in the collapse scenario ``RHD+SE C''; the four phases of evolution are as indicated.
The circle (at 2.4$\mbox{ M}_\odot$) denotes the onset of the formation of the circumstellar disk in the computational domain.
The dashed line denotes the evolution of the comparison run ``SE C'' assuming a constant accretion rate of $7\times10^{-4}  \mbox{ M}_\odot \mbox{ yr}^{-1}$.
}
\label{fig:C-Rstar_vs_Mstar}
\end{center}
\end{figure}

The protostellar evolution in the framework of the three scenarios is shown in the corresponding Hertzsprung-Russell diagrams and the evolution of the stellar radius with mass in Figs.~\ref{fig:A-HRD} to \ref{fig:C-Rstar_vs_Mstar}.  
During phase (I) before the onset of bloating, the accretion timescale is much shorter (up to four orders of magnitude) than the Kelvin-Helmholtz contraction timescale of the newborn protostar. 
Energy transport within the protostar is initially primarily via convection, but the protostar later becomes radiative.  
In either case the protostar is not in thermal equilibrium and it eventually begins to expand rapidly when the Kelvin-Helmholtz timescale is longer than the accretion timescale.
The maximum stellar radius and the mass range of the bloating phase (II) are given in Table~\ref{tab:runs}.  
When the Kelvin-Helmholtz timescale becomes shorter than the accretion timescale (phase III), the protostar begins to contract toward the main sequence. 
This contraction is stopped when hydrogen burning begins (phase IV).

\begin{figure}[htbp]
\begin{center}
\includegraphics[width=0.45\textwidth]{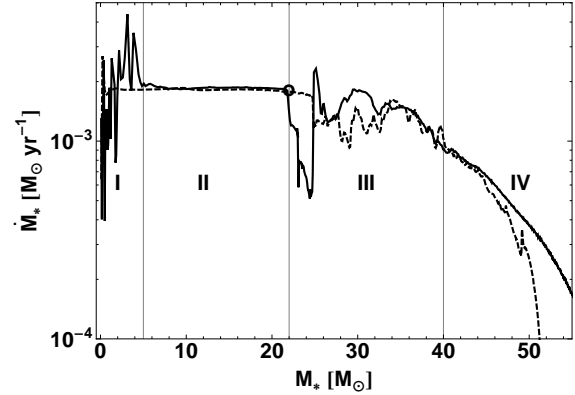}\\
\caption{
Evolution of the accretion rate as a function of the protostellar mass in the collapse scenario ``RHD+SE A'' (solid line); 
the four phases of evolution are as indicated.
The dashed line denotes the evolution of the accretion rate in a comparison run (``RHD A'') using an a priori computed stellar evolutionary track.
The circle (at 22$\mbox{ M}_\odot$) denotes the onset of the formation of the circumstellar disk in the computational domain.
}
\label{fig:A-Mdot_vs_Mstar}
\end{center}
\end{figure}

\begin{figure}[htbp]
\begin{center}
\includegraphics[width=0.45\textwidth]{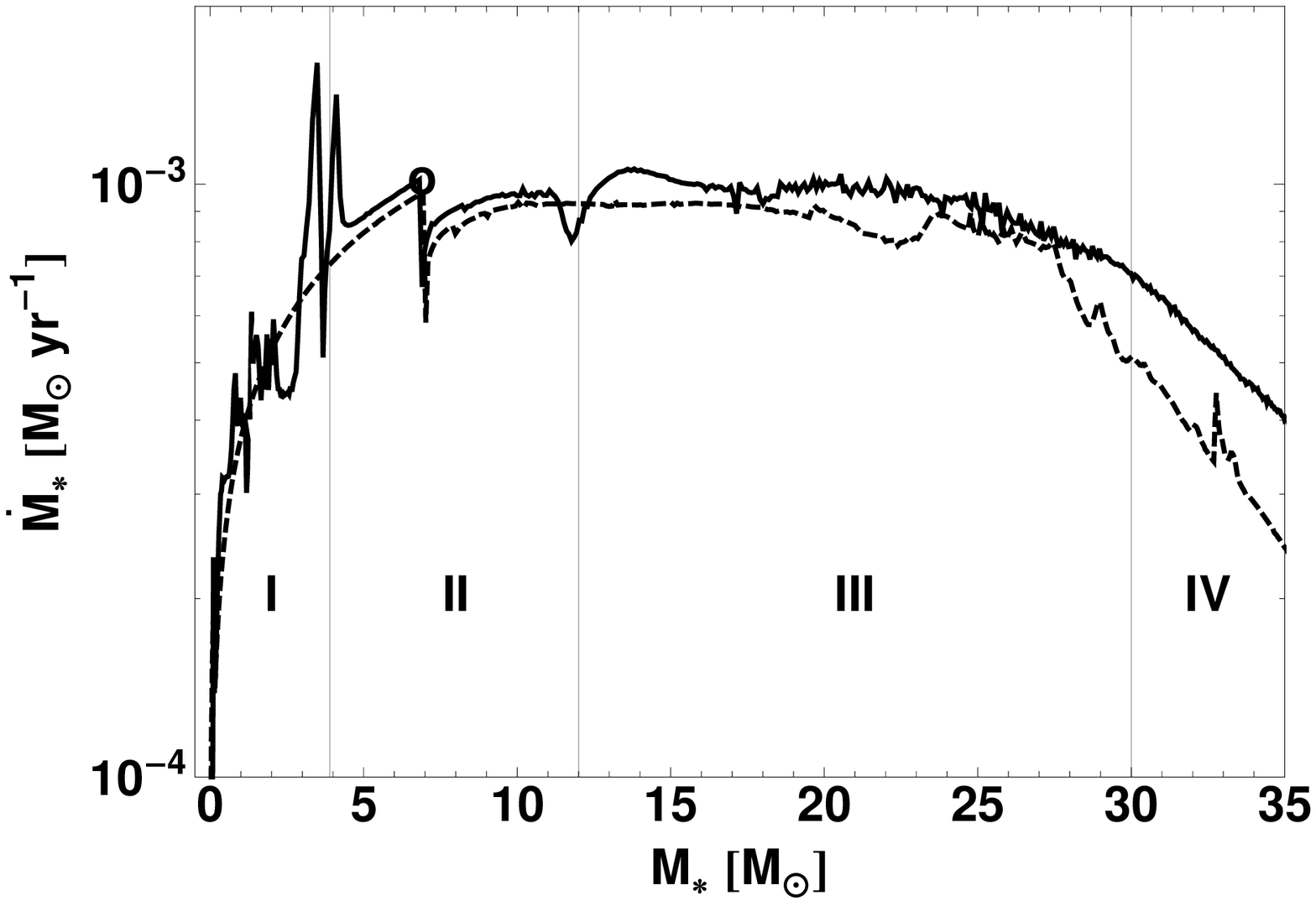}\\
\caption{
Evolution of the accretion rate as a function of the protostellar mass in the collapse scenario ``RHD+SE B''; 
the four phases of evolution are as indicated.
The dashed line denotes the evolution of the accretion rate in a comparison run (``RHD B'') using an a priori computed stellar evolutionary track.
The circle (at 7$\mbox{ M}_\odot$) denotes the onset of the formation of the circumstellar disk in the computational domain.
}
\label{fig:B-Mdot_vs_Mstar}
\end{center}
\end{figure}

\begin{figure}[htbp]
\begin{center}
\includegraphics[width=0.45\textwidth]{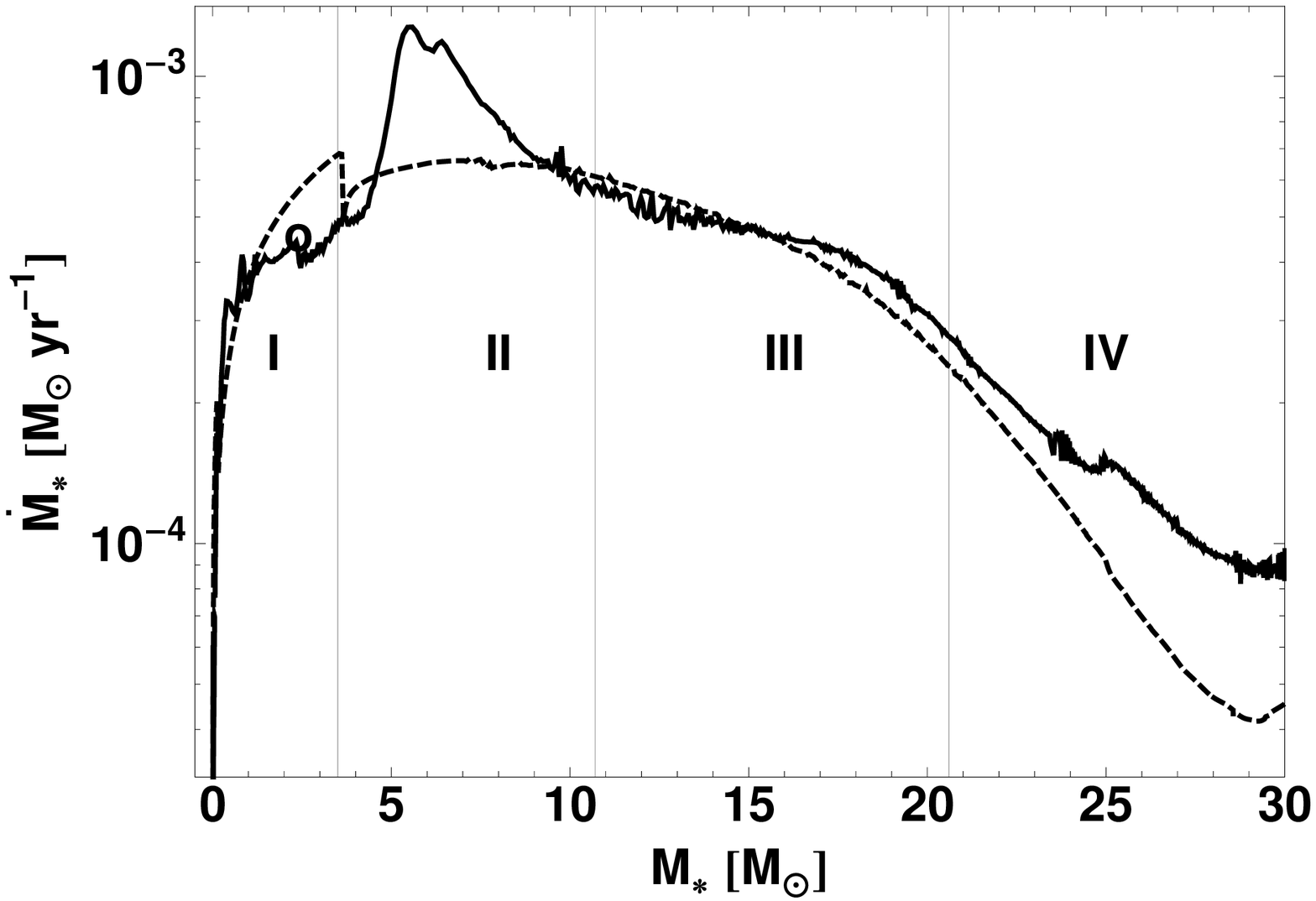}\\
\caption{
Evolution of the accretion rate as a function of the protostellar mass in the collapse scenario ``RHD+SE C''; 
the four phases of evolution are as indicated.
The dashed line denotes the evolution of the accretion rate in a comparison run (``RHD C'') using an a priori computed stellar evolutionary track.
The circle (at 2.4$\mbox{ M}_\odot$) denotes the onset of the formation of the circumstellar disk in the computational domain.
}
\label{fig:C-Mdot_vs_Mstar}
\end{center}
\end{figure}

The effects of protostellar feedback on the pre-stellar core evolution in the three scenarios is most clearly shown by the evolution of the accretion rate onto the protostar in Figs.~\ref{fig:A-Mdot_vs_Mstar} to \ref{fig:C-Mdot_vs_Mstar}. 
In general, the interdependence of protostellar and the pre-stellar core evolution results in bursts of accretion and/or a non-constant, time-varying accretion rate.
The formation of a circumstellar disk is accompanied by a sudden drop in the accretion rate followed by a short phase of enhanced accretion, although this effect is barely noticeable for case ``RHD+SE B'' (see Fig. \ref{fig:B-Mdot_vs_Mstar}), as the disk forms during phase (II).


\section{Discussion}
The results presented in the previous section demonstrate the interdependency of protostellar evolution and the evolution of its host molecular core.
The three collapse scenarios considered differ in their density slopes and rotational to gravitational energy ratios, see Table~\ref{tab:runs}.
The differing initial angular momentum distributions result in circumstellar disk formation at different phases of the protostellar evolution.
In run ``A'' the circumstellar disk forms at the end of protostellar bloating and the start of phase (III).
In run ``B'' the circumstellar disk forms during the protostellar bloating phase (II).
In run ``C'' the circumstellar disk forms prior to the protostellar bloating during phase (I).

\subsection{Phase I: Before the bloating}
\begin{figure}[htbp]
\begin{center}
\includegraphics[width=0.45\textwidth]{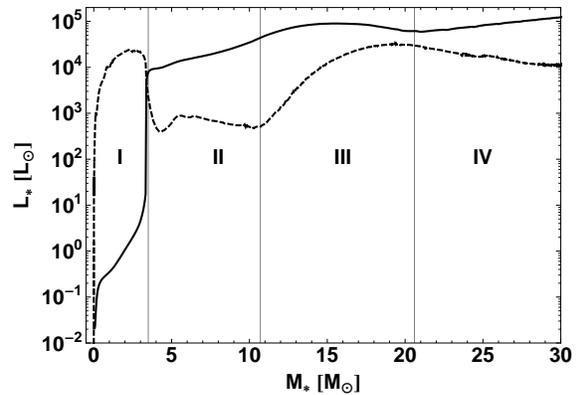}\\
\caption{
Accretion luminosity (dashed line) and protostellar luminosity (solid line) as a function of the protostellar mass in the collapse scenario ``RHD+SE C''.
}
\label{fig:C-Lstar_vs_M}
\end{center}
\end{figure}

One characteristic of the first phase is the domination of accretion luminosity over the intrinsic stellar luminosity, even for the collapse scenario ``C'' with the lowest initial accretion rate, see Fig.~\ref{fig:C-Lstar_vs_M}.
This dominant role of the accretion luminosity extends throughout phase I, which ends when the protostellar radius dramatically increases, lowering the specific energy of the material $\eta GM_* / R_*$ being accreted.  
Prior to bloating the radiative feedback of the centrally forming protostar onto the accretion flow depends strongly on the protostellar radius $R_*$ and the accretion rate $\dot{M}_*$ itself, see eq.~\eqref{eq-acc}.
This yields a negative feedback mechanism with a time delay and results in quasi-periodic oscillations of the accretion rate:
An increase of accretion at the surface of the star leads to a higher accretion luminosity, which in turn heats the nearby (dusty) protostellar environment.
Radiative acceleration on the infalling dusty gas slows the accretion flow, resulting in a decrease of the accretion rate.
After a delay given by the length of time the material needs to flow from the dust sublimation radius onto the surface of the star, the accretion luminosity decreases. 
This in turn reduced the radiative acceleration and allows the accretion rate to increase again.
Altogether, this negative feedback coupling leads to a variability of the accretion rate in all three collapse scenarios investigated.
In run ``RHS+SE C'', this general behavior is modified by the early formation of the circumstellar disk, which lowers the direct interdependence between accretion rate and accretion luminosity. 
The material to be accreted by the star now has to pass through the circumstellar disk.
Variations in the rate that the disk accretes material does not necessarily translate into analogous variations of the rate that the disk transports material to the stellar surface.
The result is lower variability of accretion luminosity and accretion rate and a smoother transition into the bloating phase.

\subsection{Phase II: Protostellar bloating phase}
In all cases considered, accreting protostars pass through the ``bloating phase'' (II). 
The maximum stellar radius of the protostar strongly depends on the accretion history of the protostar; it ranges from 
$389 \mbox{ R}_\odot$ (run ``RHD+SE C'') 
up to 
$885 \mbox{ R}_\odot$ (run ``RHD+SE A'').  
Similar values are obtained at constant average accretion rates of 
$7 \times 10^{-4} \mbox{ M}_\odot \mbox{ yr}^{-1}$ (run ``SE C'')
to 
$2 \times 10^{-3} \mbox{ M}_\odot \mbox{ yr}^{-1}$ (run ``SE A''), respectively.

During the bloating phase there is little interdependency between the stellar and core evolution.
The large stellar radius implies a significant decrease of the Kelvin-Helmholtz contraction timescale 
of the protostar; in fact it is even smaller than the accretion timescale and the protostar's evolution is not significantly affected by small variations of the accretion rate.
Moreover, the stellar luminosity is shifted to the infrared during the bloating phase, which decreases the radiative interaction with the (nearby) accretion flow and the molecular core's evolution is less dependent on the evolution of the protostar. 
As a result, the simulation run ``RHD+SE B'', in which the circumstellar disk forms during the bloating phase, shows the smallest differences when compared to run ``RHD B''.
Furthermore, the independence of the stellar evolution on the accretion rate during the bloating phase yields a very good agreement of the accretion rate during the bloating phase in run ``RHD+SE A'' and its comparison run ``RHD A''.
In run ``RHD+SE C'', this general behavior is superimposed by the prior formation of the circumstellar disk during the ``phase I''.
The lower accretion rate from the disk to the star at the end of ``phase I'' yields an accumulation of the mass in the accretion disk, which is liberated during the bloating phase (``phase II'') due to the decrease in the accretion luminosity feedback.

\subsection{Phase III: From the bloating phase to the zero-age-main-sequence}
\begin{figure}[htbp]
\begin{center}
\includegraphics[width=0.45\textwidth]{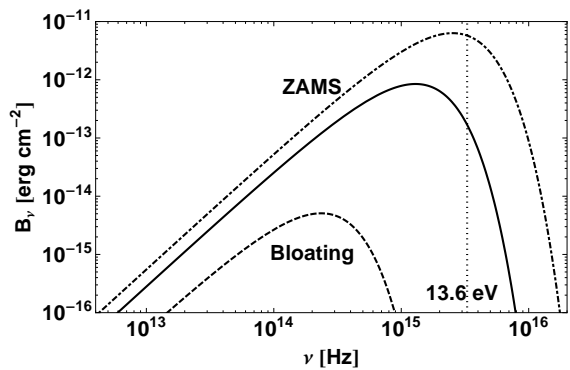}\\
\caption{
Planck spectra of the protostellar effective surface temperature 
at the end of the bloating phase (dashed line), 
after the initial stellar contraction and disk formation (solid line), and 
at the beginning of the ZAMS (dot-dashed line) 
in the collapse scenario ``RHD+SE A''.
The vertical dotted line denotes the hydrogen ionization threshold of 13.6~eV.
}
\label{fig:A-Planck}
\end{center}
\end{figure}

After the bloating phase, the high mass of the protostar yields a stellar luminosity, which continues to dominate over the accretion luminosity, even for collapse scenario ``A'' with the highest accretion rates.
This transitional phase is defined by the decrease of the stellar radius from the largest stellar radius at the end of the bloating phase to a local minimum, which denotes the zero-age-main-sequence.
In this phase, the strongest interdependency of the stellar and the core evolution is observed in run ``RHD+SE A''.
The reason is that in the collapse scenario ``A'' the circumstellar disk forms directly at the end of the protostellar bloating phase.
The disk formation yields a decrease of the accretion rate.
Due to the diminished supply of accreting gas mass, the protostar contracts much faster (factor of two) than in the comparison stellar evolution run (``SE A'') at constant accretion.
The strong decrease in the protostellar radius yields a rapid shift of the stellar spectrum towards the EUV regime, see Fig.~\ref{fig:A-Planck}.
The exerted radiation pressure of this event results in a strong decrease in the accretion rate with respect to the core evolution comparison run ``RHD A''.
Ultimately, the protostar reaches the ZAMS at $40 \mbox{ M}_\odot$.
\vONE{
Furthermore, the rapid shift of the stellar spectrum depicts a sudden onset of ionization.
In the radiation-hydrodynamics evolution of the stellar environment, the effect of ionization is not considered here.
For the effects of ionization feedback during the pre-main-sequence evolution of massive stars and implications on observations of H II region variability, we refer the interested reader to recent studies by \citet{Peters:2010p16, Peters:2010p783}, and \citet{Klassen:2012p19028}.
}

\subsection{Phase IV: Main sequence evolution}
In the various collapse scenarios, the protostars reach the ZAMS at $21 \mbox{ M}_\odot$, $30 \mbox{ M}_\odot$, and $40 \mbox{ M}_\odot$, respectively.
The high stellar masses correspond to lower Kelvin-Helmholtz contraction timescales, leading in general to a decreased dependency of the stellar evolution on the accretion rate.
After reaching the ZAMS, the protostellar radius grows along the ZAMS as its mass increases.
We stopped the simulation runs shortly after the (proto)star has reached the ZAMS.

\section{Aims and limitations}
The stellar evolutionary tracks shown here are in reasonably good agreement with previous studies on the evolution of massive protostars including the effect of accretion by \citet{Hosokawa:2009p23} and \citet{Hosokawa:2010p690}.
The initial dominant role of the accretion luminosity, the bloating phase of the protostar, and a smooth transition from the bloating phase towards the ZAMS are robust features in all these studies.  
However, since the evolution of the protostar will -- in general -- depend on the initial starting model, we find some differences in the early evolution to the published results by \citet{Hosokawa:2010p690}.
As an example, the maximum stellar radius during the bloating phase computed here more closely resembles the result by \citet{Hosokawa:2010p690} for their initial condition of a shallower initial entropy distribution (compare Fig.~9 in \citet{Hosokawa:2010p690}), although they assume a much larger initial stellar radius for their starting models.
%\subsection{Boundary condition}
Moreover, the evolution of the protostar depends on the outer boundary condition as well.
The outer boundary condition of the stellar evolutionary model is determined by the detailed geometry of the accretion flow onto the stellar surface and how much heat is deposited into the outer stellar layers.
In this study, the so-called ``cold disk accretion'' boundary condition is used, in which the kinetic energy from accretion is assumed to be radiated away without significantly penetrating the stellar atmosphere.
%For further details and a comparison study of different boundary conditions, please see \citet{Hosokawa:2010p690}.

\begin{figure}[htbp]
\begin{center}
\subfigure{\includegraphics[width=0.45\textwidth]{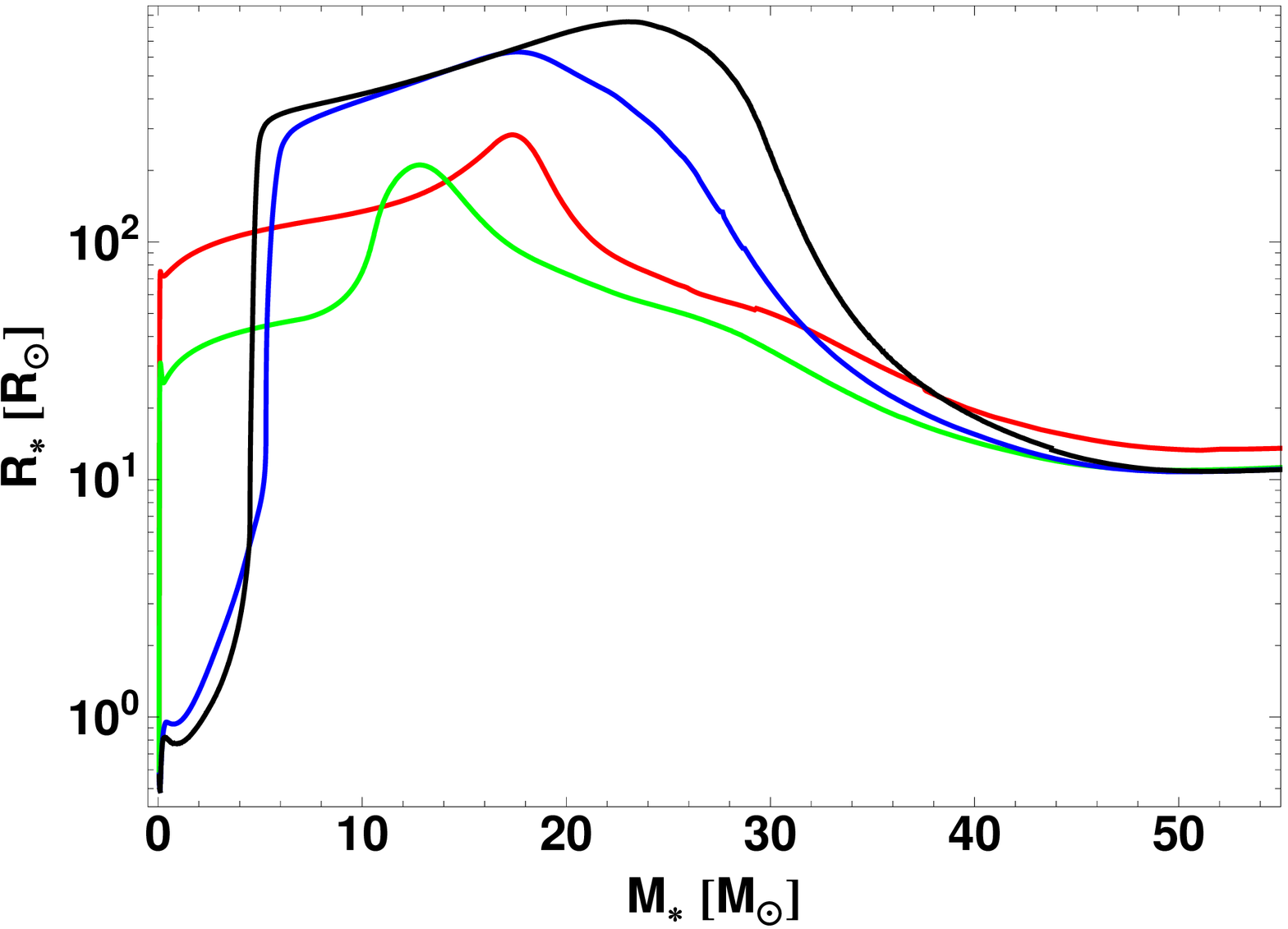}}
\subfigure{\includegraphics[width=0.45\textwidth]{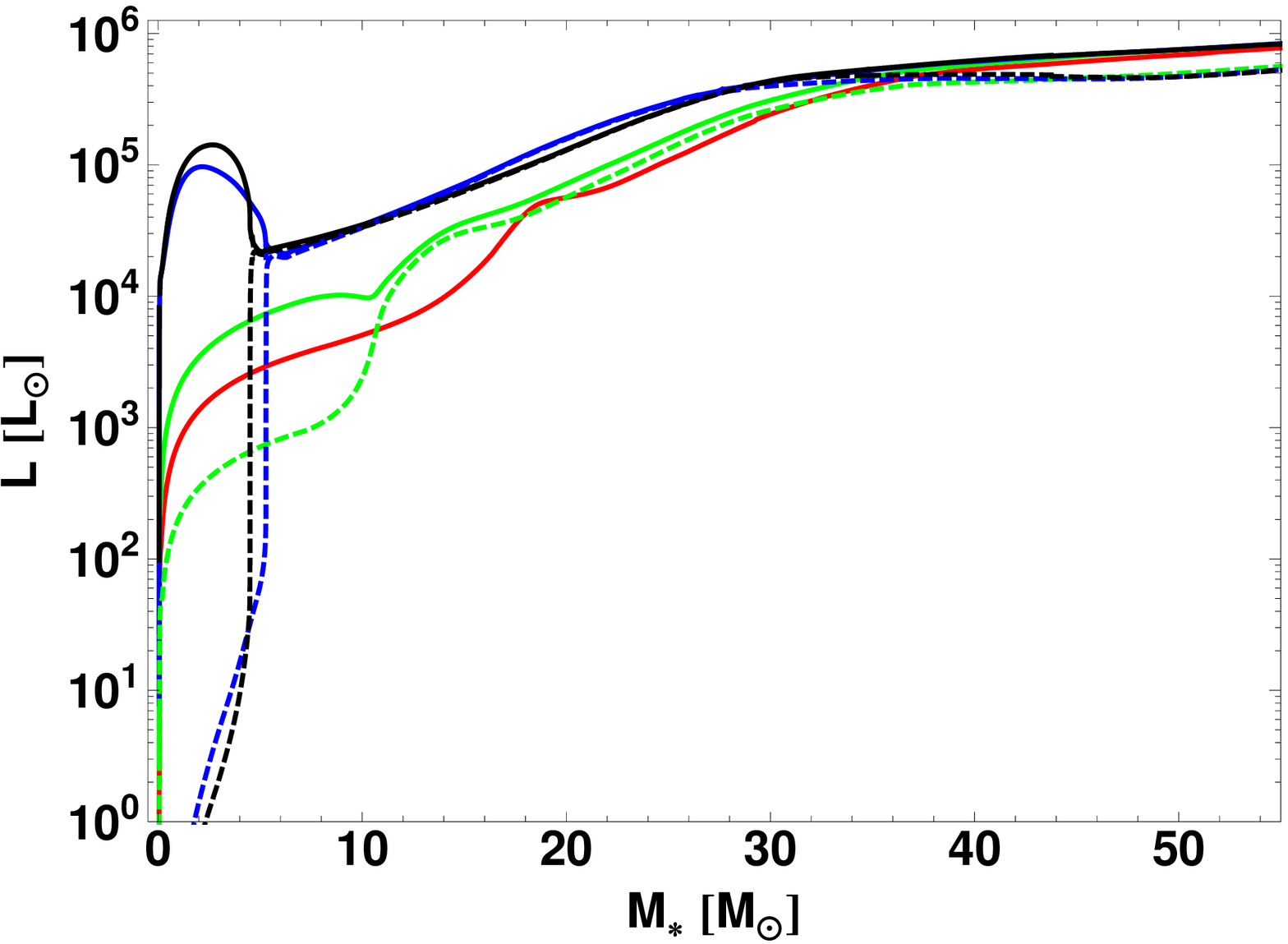}}
\caption{
Protostellar radius (upper panel)
and
luminosities (lower panel) 
as function of stellar mass for different absorption efficiencies $\eta$ of the accretion energy. 
The black lines denote the limiting case of $\eta=0$, a so-called cold disk accretion model, in which the kinetic energy from accretion is assumed to be radiated away without significantly penetrating the stellar atmosphere.
The blue lines denote the case of $\eta=10^{-6}$.
The green lines denote the case of $\eta=0.1$.
The red lines denote the limiting case of $\eta=1$, in which the total luminosity of the accreting protostar is just given by its intrinsic stellar luminosity and the accretion energy is fully deposited into the protostellar atmosphere.
In the lower panel, solid lines denote the total luminosity including the accretion luminosity and dashed lines denote the intrinsic stellar luminosity only.
}
\label{fig:eta}
\end{center}
\end{figure}

%\vONE{
In detail, the protostellar evolution depends on the actual physics of this accretion layer.
As an example, we have additionally computed the protostellar evolution for an accretion rate of $\dot{M} = 2 \times 10^{-3} \mbox{ M}_\odot \mbox{ yr}^{-1}$ for cases, in which a fraction of the kinetic accretion energy is absorbed by the protostellar atmospheric layer.
In fact, we have studied the limiting and intermediate cases of
cold disk accretion ($\eta = 0$, the preset model for the previous RHD+SE and SE runs),
minor energy absorption ($\eta = 10^{-6}$),
significant energy absorption ($\eta = 0.1$), and
full energy absorption ($\eta = 1$).
The corresponding evolution of the stellar radii and luminosities for these accretion layer scenarios are shown in Fig.~\ref{fig:eta}.
% and \ref{fig:eta2}.

%\begin{figure}[htbp]
%\begin{center}
%\includegraphics[width=0.45\textwidth]{./fig13.eps}\\
%\caption{
%Luminosities as function of stellar mass for different absorption efficiencies $\eta$ of the accretion energy. 
%Black lines denote the total luminosity including the accretion luminosity.
%Red lines denote the stellar luminosity only.
%The solid line denotes the limiting case of $\eta=0$, a so-called cold disk accretion model, in which the kinetic energy from accretion is assumed to be radiated away without significantly penetrating the stellar atmosphere.
%The dashed line denotes the case of $\eta=10^{-6}$.
%The dotted line denotes the case of $\eta=0.1$.
%The dot-dashed line denotes the limiting case of $\eta=1$, in which the total luminosity of the accreting protostar is just given by its stellar luminosity and the accretion energy is fully deposited into the protostellar atmosphere.
%}
%\label{fig:eta2}
%\end{center}
%\end{figure}
In this case of an extremely high accretion rate, the maximum radius of the bloated protostar depends quite strongly on the absorption efficiency $\eta$.
While the run of minor energy absorption ($\eta = 10^{-6}$) closely resembles the case of cold disk accretion ($\eta = 0$), the models for significant ($\eta = 0.1$) and full deposition of the accretion energy ($\eta = 1$) yield significantly larger radii during the pre-bloated phase and smaller radii during the bloated phase.
Eventually, the models merge when the protostar reaches the zero age main sequence.
%}

Our particular choice of the initial model and boundary conditions for the stellar evolution code
were not made specifically to quantitatively reproduce any particular previous stellar evolution result.
The main goal of this study is to illuminate the interdependencies of protostellar evolution and the evolution of the host molecular core.
Thus, to allow comparisons between our self-consistent protostellar evolutionary tracks presented here in three different collapse scenarios, we do not vary the stellar evolution parameters, leaving this type of parameter study for a future investigation.
\vONE{
%Fragmentation
Regarding the hydrodynamical evolution of the proto-stellar environment, we point out that in these axially symmetric studies fragmentation of the forming accretion disk is neglected.
%Nonetheless, this study focusses on the proto-stellar evolutionary phase only and disk fragmentation is expected to occur later in time.
}
%\vTWO{
%}

\section{Observing bloated massive protostars}
Apart from their long distance, observations of massive protostars in general suffer from the fact that the protostars are -- especially at these early time in evolution -- deeply embedded into their natal environment.
On the other hand, due to their high luminosity in combination with a low surface temperature during the bloating phase these massive protostars will have unique signatures.
First signs of a strongly bloated, relatively cool central object were reported in an interferometry measurement by \citet{Linz:2009p3056}.
However, this represents an indirect detection due to the fact that the conclusion of a bloated cool object was obtained via a best fit model to the observed data.
A second candidate for a massive protostar close to the end of its bloating phase is discussed in \citep{Bik:2012p3198}.

To alleviate the extinction problem the best observational chances are given for cases, in which e.g.~a protostellar outflow has formed bipolar cavities of low density, which are aligned close to the line of sight.
If we assume that the outflow launching is connected to the existence of an accretion disk, this corresponds to an early disk formation scenario such as the collapse scenario ``C'' herein.
The luminosity of the protostar can then be determined by its far infrared spectrum.
Measuring the stellar surface temperature is potentially achievable via very high signal to noise ratio K-band spectroscopy. 
The stellar absorption lines will be heavily veiled by the emission of the circumstellar disk and envelope.
These bloated stars will look like red super giants and absorption lines of Mg I and Na I  and CO will be observable \citep{Rayner:2009p4328}.
Another concern arises, if we have in mind that the hot inner rim of the accretion disk can have temperatures comparable to the one of the bloated protostar.
Disentangling both temperatures by rotational signatures of the disk material is mostly ruled out due to the fact that we assume to observe the protostar face-on through a low density cavity, forming perpendicular to the accretion disk.

An alternative technique to the one described above is to measure the scattered light in the cavity walls, in case they are not aligned with the line of sight \citep{Testi:2010p4301}.

\section{Summary and Outlook}
We have studied the simultaneous evolution of massive protostars within their evolving host molecular cores.
The evolution of the stellar environment is computed with a radiation hydrodynamics code that includes self-gravity and thermal and radiative acceleration feedback.
The evolution of the protostar is computed with a stellar evolution code that includes the effects of accretion.
The interdependencies of the evolving protostar and molecular core emerge in a variety of phenomena 
and are studied in three examples of different collapse scenarios:
The circumstellar accretion disk forms either prior (collapse scenario ``C''), during (collapse scenario ``B''), or at the end (collapse scenario ``A'') of the protostellar bloating phase.

The protostellar evolutionary tracks calculated here show the robust features of massive protostellar evolution at high accretion rates:
Initially, the accretion luminosity dominates over the stellar luminosity.
The importance of accretion luminosity decreases with the onset of the bloating phase; 
the protostar's radius approaches values up to several $100~\mbox{R}_\odot$, depending on the accretion rate.
After the bloating phase, the protostellar radius decreases until the protostar reaches the ZAMS.

The formation of a circumstellar disk strongly influences the early evolution of the central massive protostar by acting as a ``valve'' and ``low pass filter'' for accretion onto the protostar.
The disk's impact strongly depends on the order of events, in particular when the disk is formed with respect to the bloating phase of the star, which itself depends on the initial angular momentum distribution of the local environment.
The highest impact on the protostellar evolution occurs, if the disk forms at the end of the bloating phase.
The impact is minimal, if the disk forms during the bloating phase.

Approximate evolutionary tracks can provide a reasonable estimate of the protostar's influence on
the molecular environment, if the mean accretion rate onto the protostar is known.  
In the early phases of evolution this influence is principally the radiation pressure exerted by the new-born protostar. 
Later, when the massive protostar reaches the ZAMS, the ionizing flux will ionize and heat the surroundings and can ultimately destroy the accretion disk through photoevaporation \citep{Zinnecker:2007p363}. 
Prior to reaching the main sequence, ionization of the surroundings is efficiently suppressed.

It is likely that a number of effects not considered here will contribute to the variability of the accretion rate onto the protostar.  
We have adopted an ``$\alpha$'' formalism for angular momentum transfer in the disk \citep{Shakura:1973p689}. 
A more precise treatment would necessarily include magnetic fields and the details of their coupling to circumstellar material in a fully three dimensional radiation-magneto-hydrodynamics code.  
The effects of turbulence, outflows, fragmentation, tidal forces, changes in the abundances of various chemical species and their charged state, gas and dust opacities, and clumpy accretion onto the disk could all lead to increased variability, and some of these processes could both influence and be influenced by protostellar evolution. 
In this sense, this study should be considered a first step in understanding some of the basic feedback effects of accretion and early protostellar evolution.
Further work is needed and will be the subject of future studies.

Finally, with respect to observations, the collapse scenario ``C'' depicts the most promising example case. 
The early formation of a circumstellar accretion disk will potentially lead to the launching of a protostellar outflow, which in turn will lead to the formation of low density bipolar cavities.
Observations along such cavities (or observations of scattered light at the cavity walls) could lead to the first direct detection of a massive bloated protostar.

\acknowledgments
This research project was financially supported by the German Academy of Science Leopoldina within the Leopoldina Fellowship programme, grant no.~LPDS 2011-5.
We thank our colleague Takashi Hosokawa for fruitful discussions and support.
Author R.~K.~thanks Arjan Bik, Hendrik Linz, and Henrik Beuther for discussing the observational chances and most promising techniques towards detecting massive protostars during their bloating phase.
This work was conducted at the Jet Propulsion Laboratory, California Institute of Technology, operating under a contract with the National Aeronautics and Space Administration (NASA).

\appendix
\section{Additional visualizations}
\begin{figure}[htbp]
\begin{center}
\includegraphics[width=0.45\textwidth]{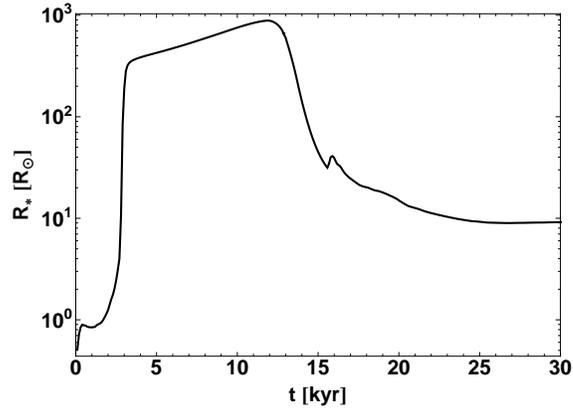}\\
\caption{
Evolution of the protostellar radius as a function of time in the collapse scenario ``RHD+SE A'' (solid line).
%The dashed line denotes the evolution of the comparison run ``SE A'' assuming a constant accretion rate of $2\times10^{-3}  \mbox{ M}_\odot \mbox{ yr}^{-1}$.
}
\label{fig:A-Rstar_vs_t}
\end{center}
\end{figure}

\begin{figure}[htbp]
\begin{center}
\includegraphics[width=0.45\textwidth]{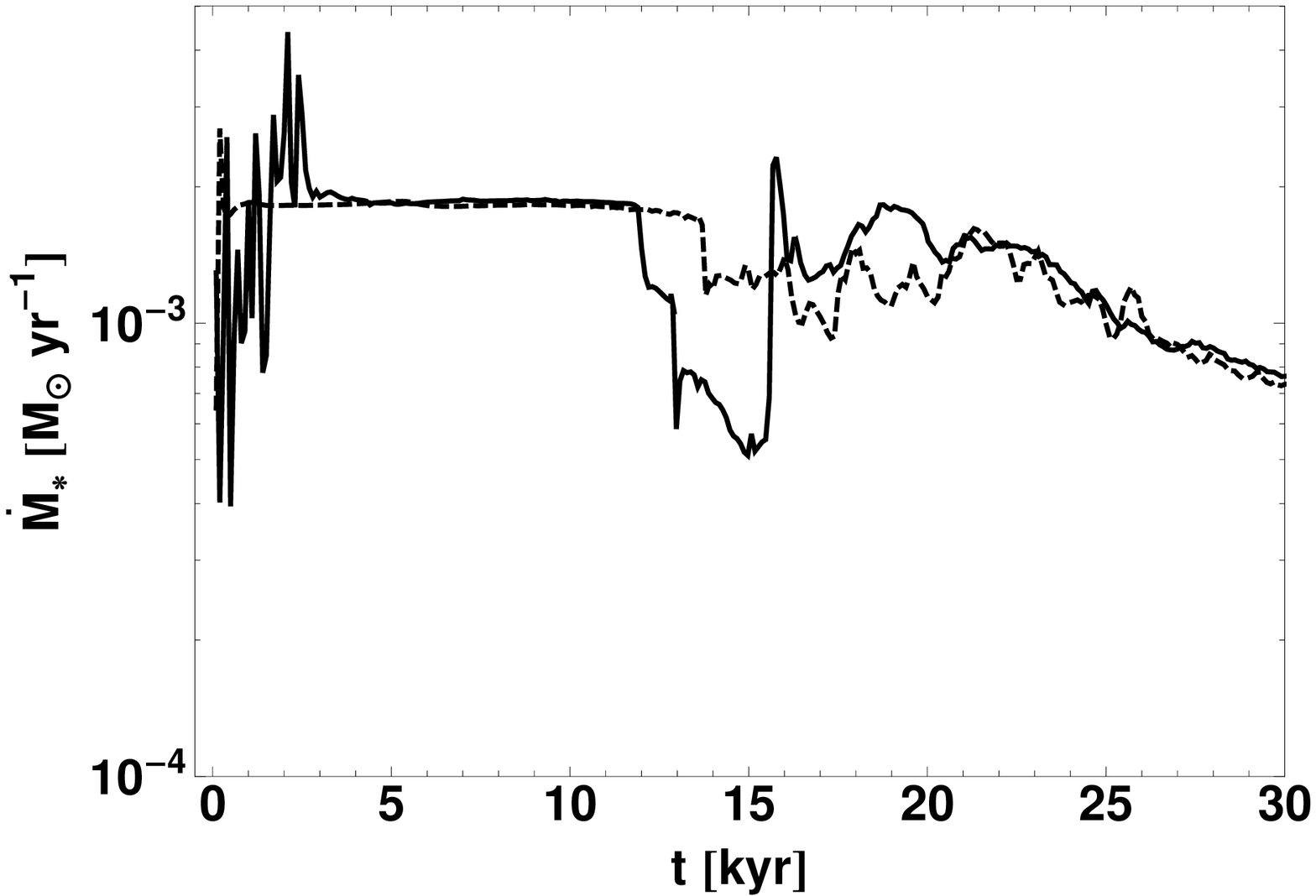}\\
\caption{
Evolution of the accretion rate as a function of time in the collapse scenario ``RHD+SE A'' (solid line).
The dashed line denotes the evolution of the accretion rate in a comparison run (``RHD A'') using an a priori computed stellar evolutionary track.
}
\label{fig:A-Mdot_vs_t}
\end{center}
\end{figure}

\begin{figure}[htbp]
\begin{center}
\includegraphics[width=0.45\textwidth]{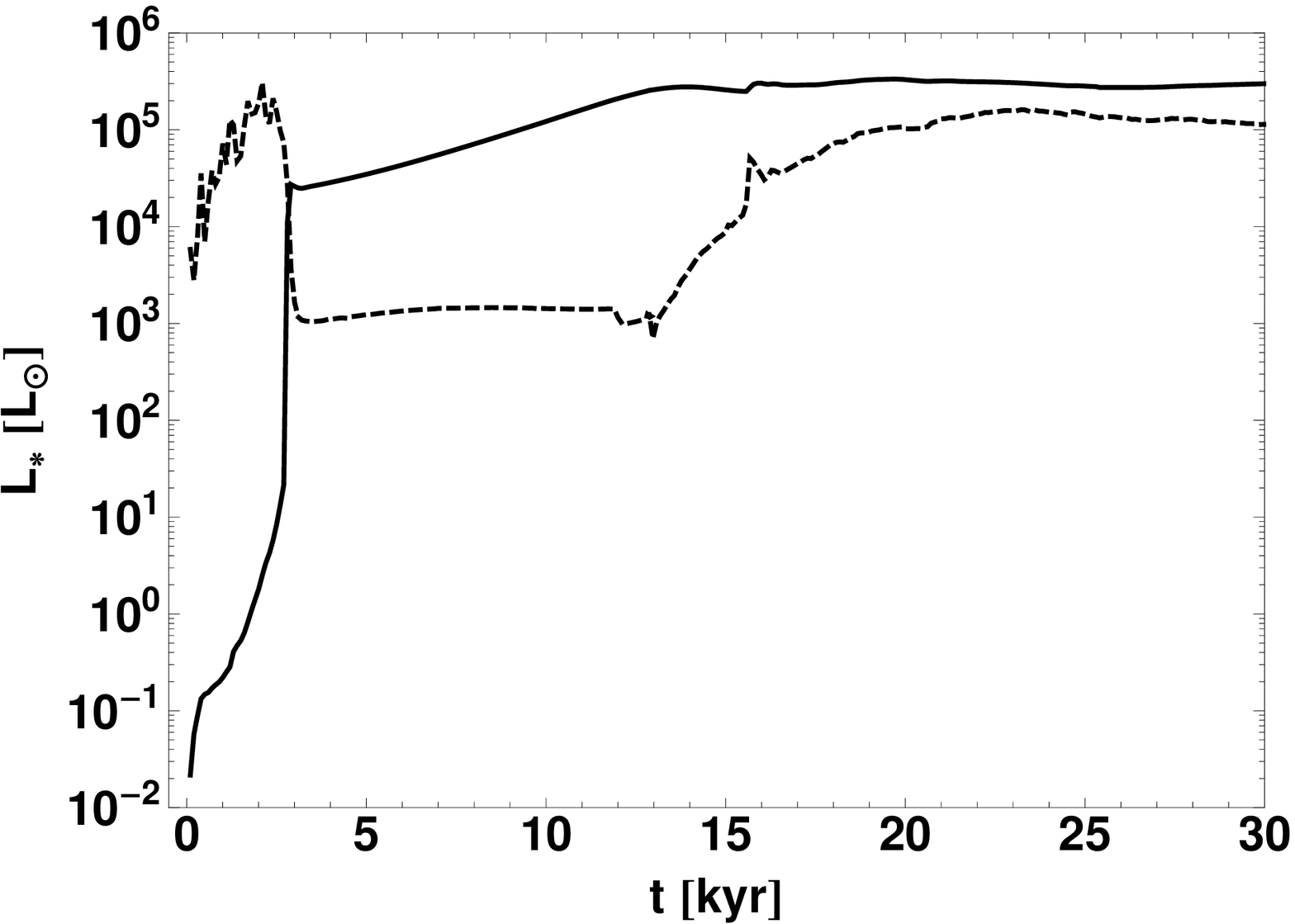}\\
\caption{
Accretion luminosity (dashed line) and protostellar luminosity (solid line) as a function of time in the collapse scenario ``RHD+SE A''.
}
\label{fig:A-Lstar_vs_t}
\end{center}
\end{figure}

\begin{figure}[htbp]
\begin{center}
\includegraphics[width=0.45\textwidth]{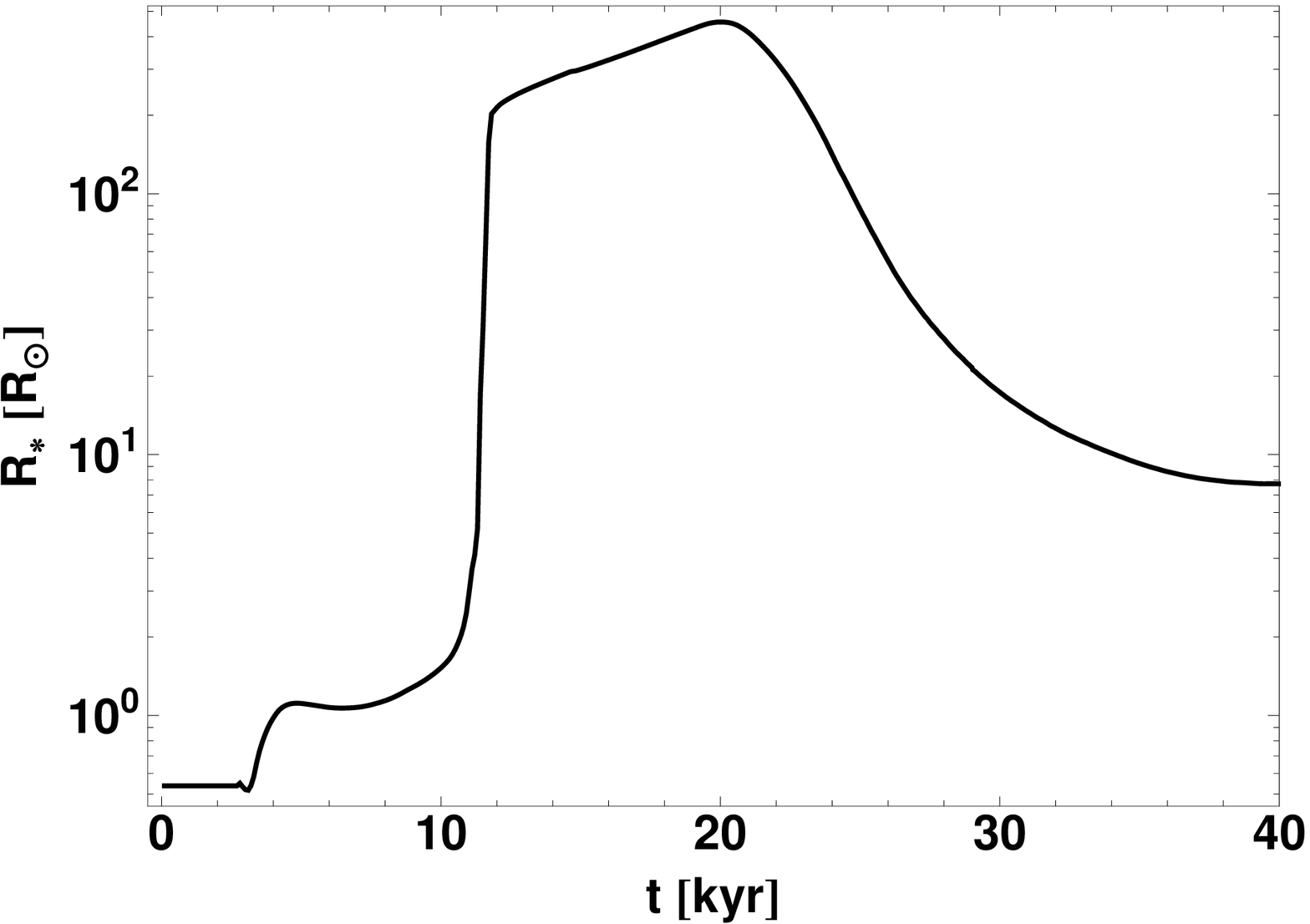}\\
\caption{
Evolution of the protostellar radius as a function of time in the collapse scenario ``RHD+SE B'' (solid line).
%The dashed line denotes the evolution of the comparison run ``SE B'' assuming a constant accretion rate of $9\times10^{-4}  \mbox{ M}_\odot \mbox{ yr}^{-1}$.
}
\label{fig:B-Rstar_vs_t}
\end{center}
\end{figure}

\begin{figure}[htbp]
\begin{center}
\includegraphics[width=0.45\textwidth]{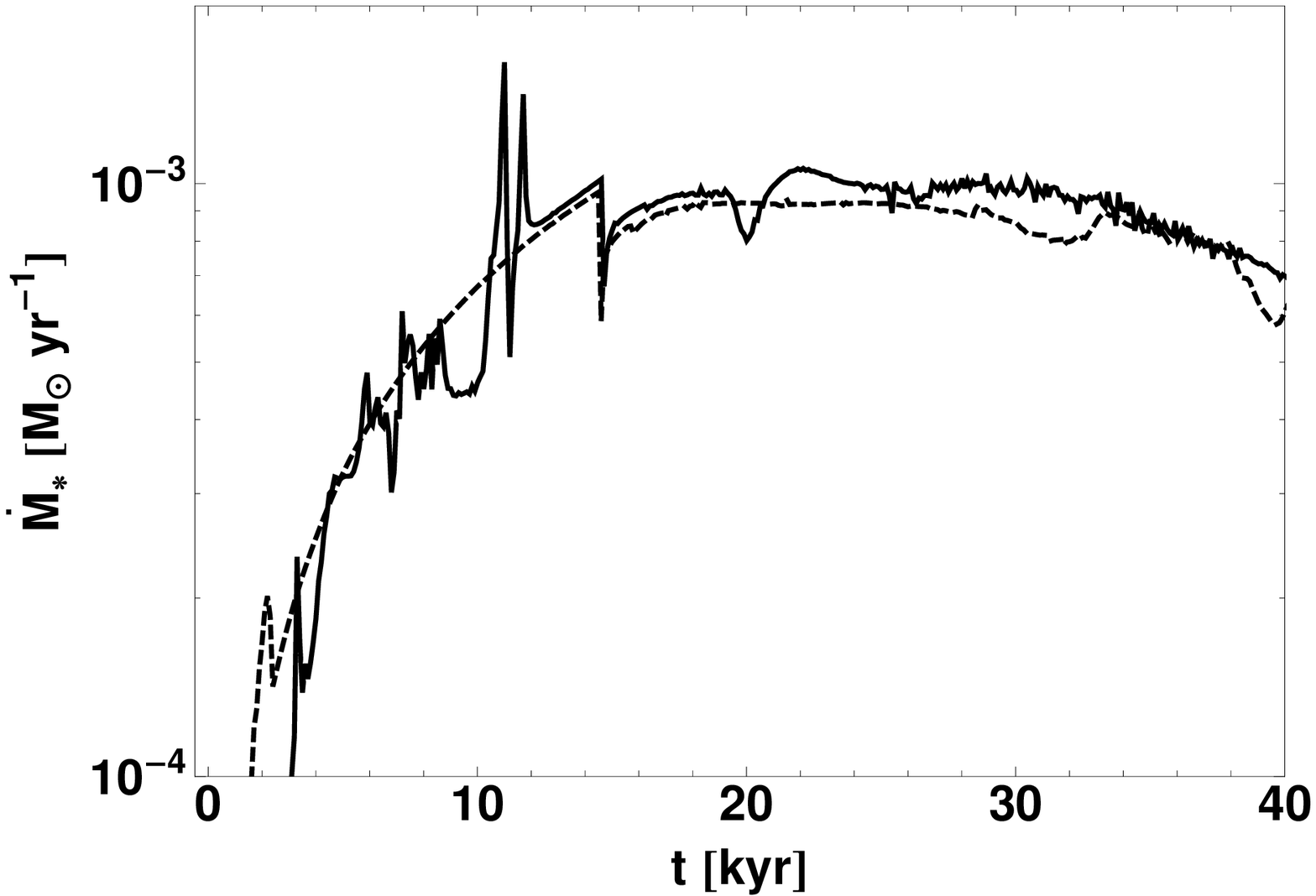}\\
\caption{
Evolution of the accretion rate as a function of time in the collapse scenario ``RHD+SE B'' (solid line).
The dashed line denotes the evolution of the accretion rate in a comparison run (``RHD B'') using an a priori computed stellar evolutionary track.
}
\label{fig:B-Mdot_vs_t}
\end{center}
\end{figure}

\begin{figure}[htbp]
\begin{center}
\includegraphics[width=0.45\textwidth]{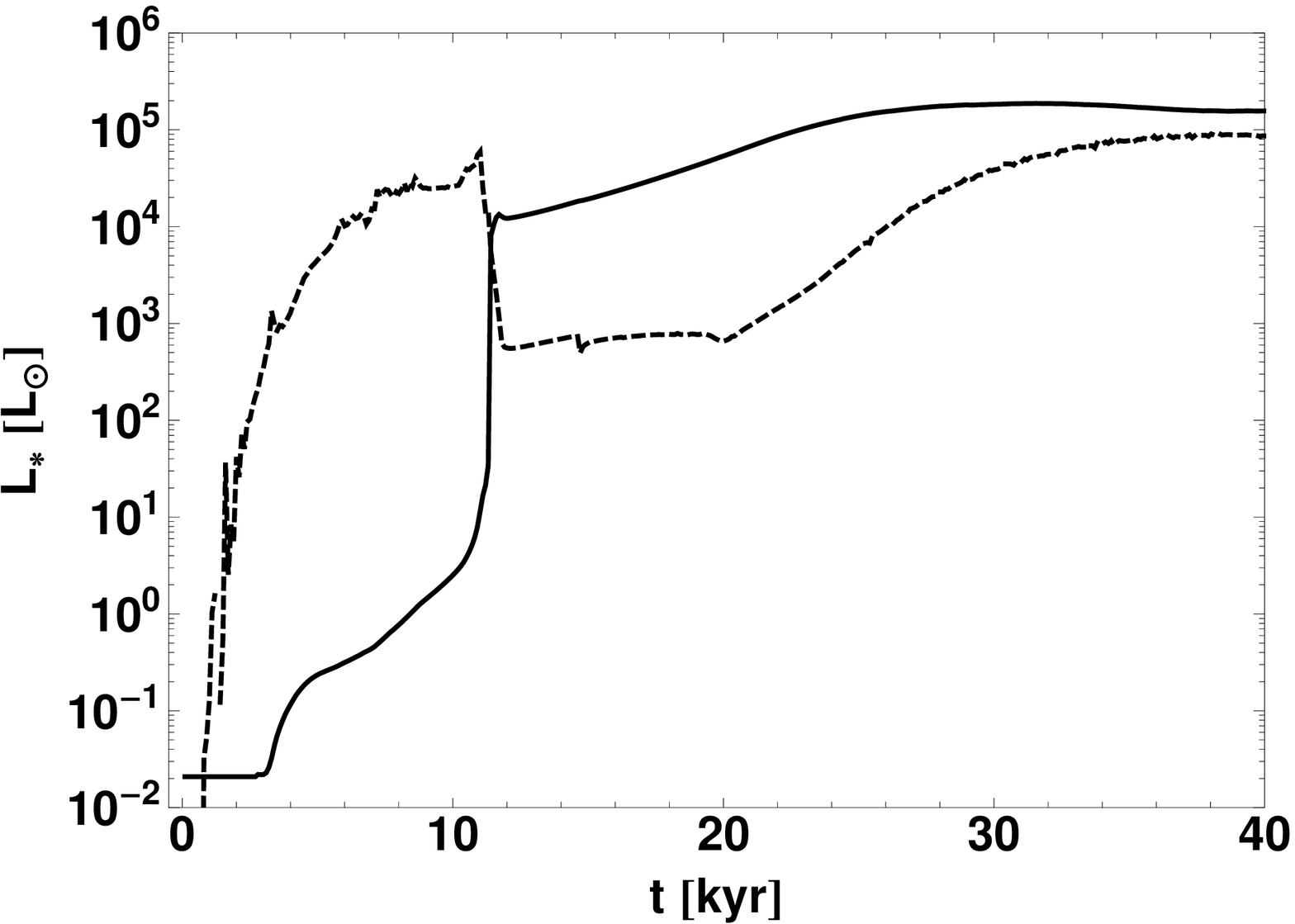}\\
\caption{
Accretion luminosity (dashed line) and protostellar luminosity (solid line) as a function of time in the collapse scenario ``RHD+SE B''.
}
\label{fig:B-Lstar_vs_t}
\end{center}
\end{figure}

\begin{figure}[htbp]
\begin{center}
\includegraphics[width=0.45\textwidth]{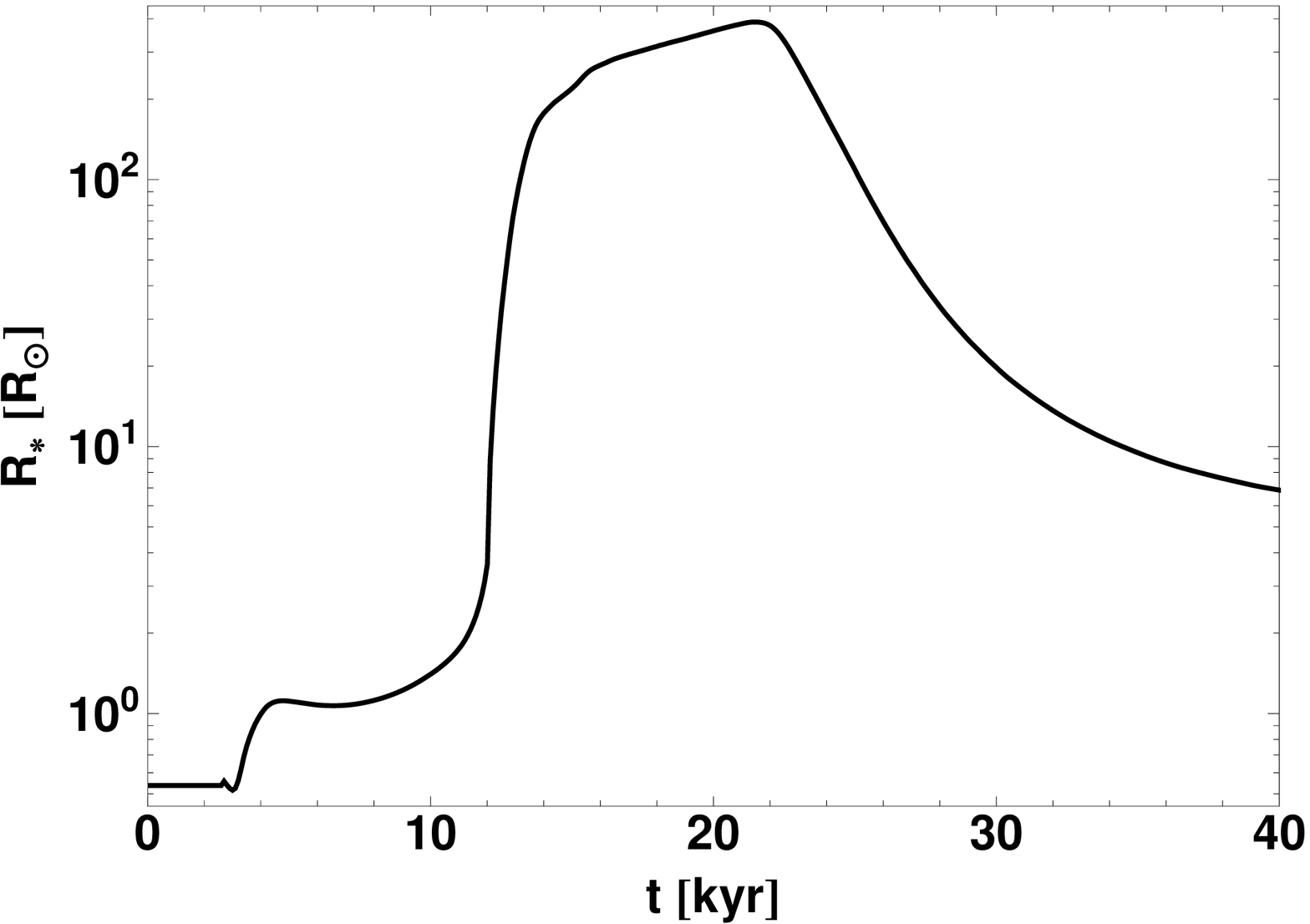}\\
\caption{
Evolution of the protostellar radius as a function of time in the collapse scenario ``RHD+SE C'' (solid line).
%The dashed line denotes the evolution of the comparison run ``SE C'' assuming a constant accretion rate of $7\times10^{-4}  \mbox{ M}_\odot \mbox{ yr}^{-1}$.
}
\label{fig:C-Rstar_vs_t}
\end{center}
\end{figure}

\begin{figure}[htbp]
\begin{center}
\includegraphics[width=0.45\textwidth]{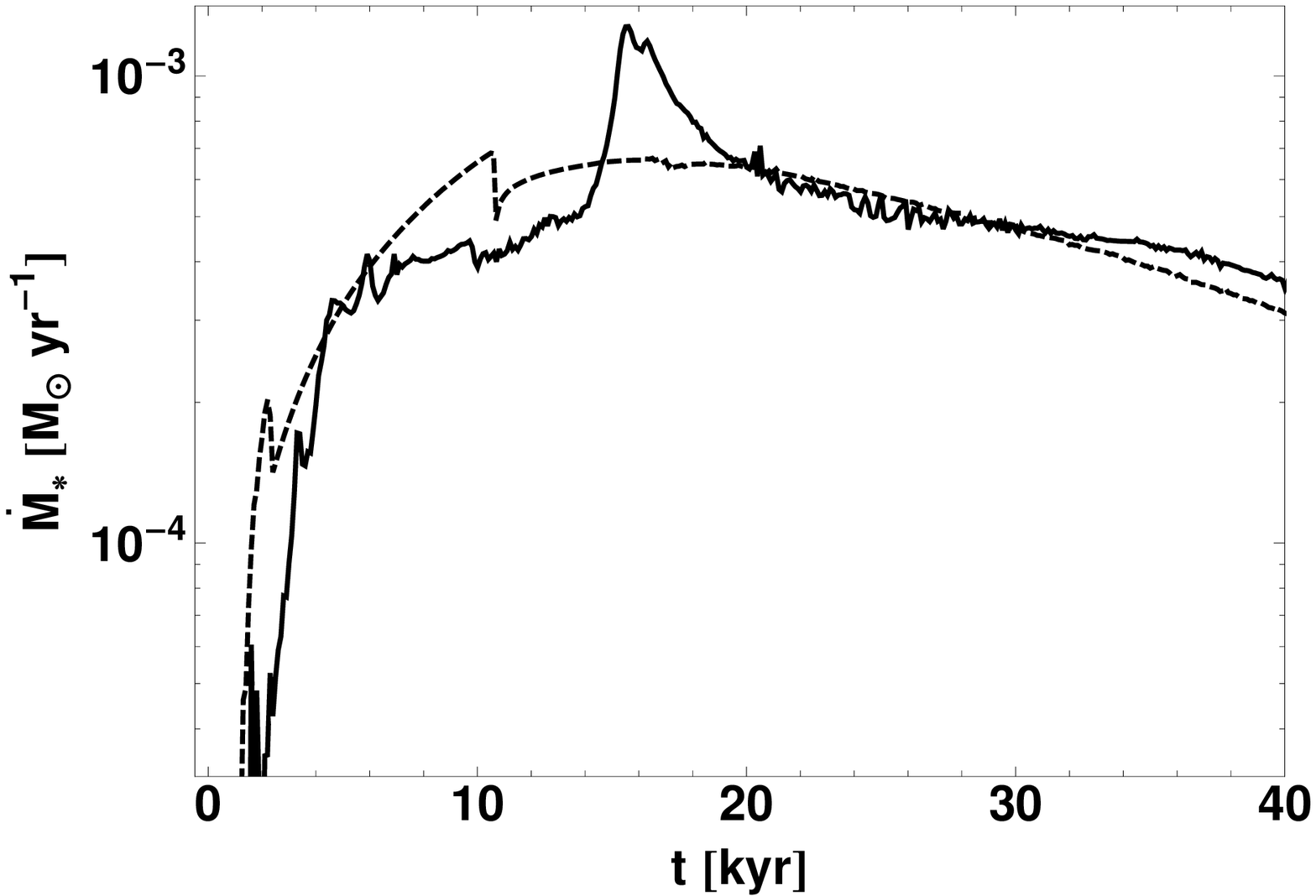}\\
\caption{
Evolution of the accretion rate as a function of time in the collapse scenario ``RHD+SE C'' (solid line).
The dashed line denotes the evolution of the accretion rate in a comparison run (``RHD C'') using an a priori computed stellar evolutionary track.
}
\label{fig:C-Mdot_vs_t}
\end{center}
\end{figure}

\begin{figure}[htbp]
\begin{center}
\includegraphics[width=0.45\textwidth]{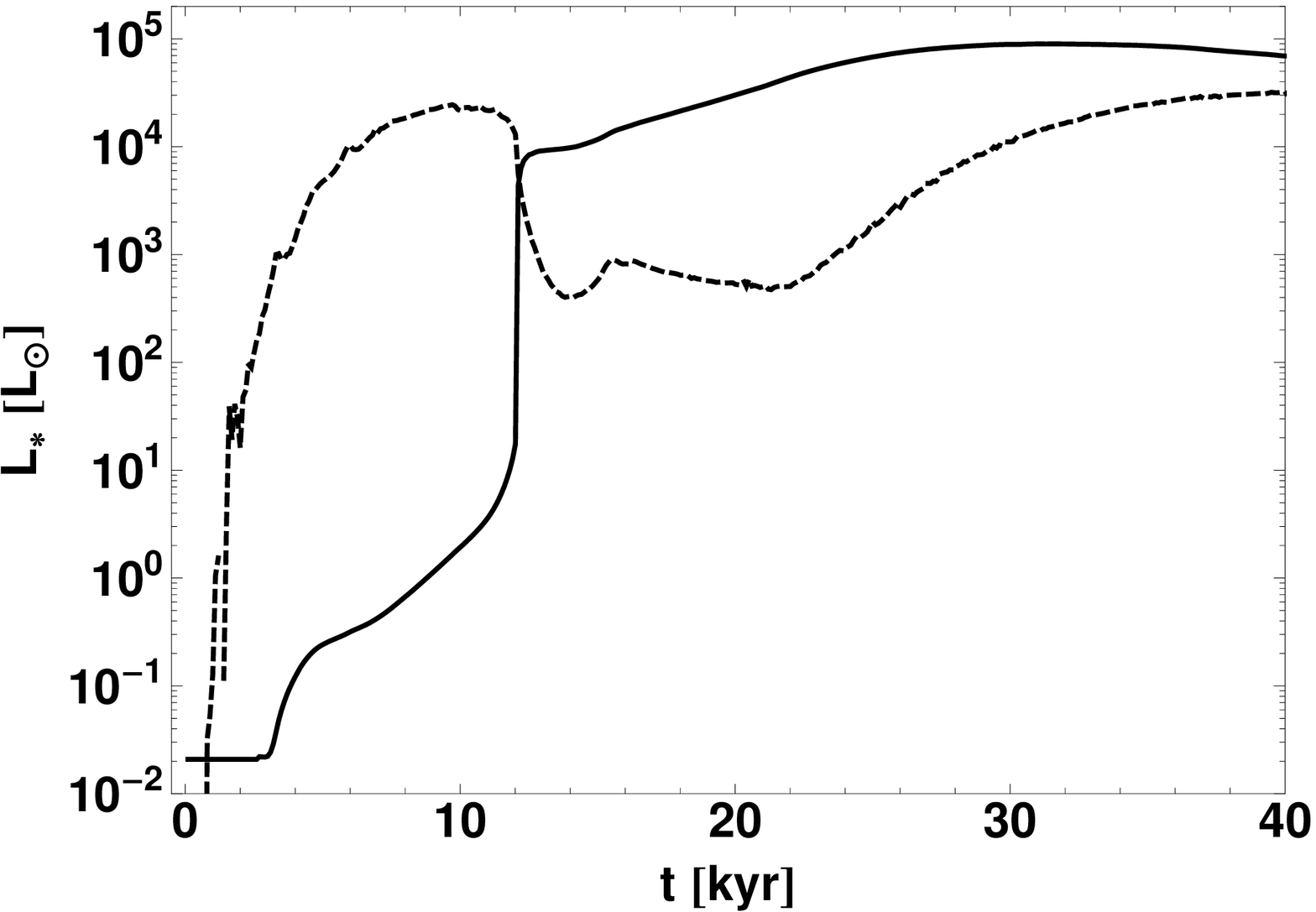}\\
\caption{
Accretion luminosity (dashed line) and protostellar luminosity (solid line) as a function of time in the collapse scenario ``RHD+SE C''.
}
\label{fig:C-Lstar_vs_t}
\end{center}
\end{figure}

For completeness,
we additionally visualize the evolution of the important variables ($R(t), \dot{M}(t)$, and L(t)) with time in Figs.~\ref{fig:A-Rstar_vs_t} to \ref{fig:C-Lstar_vs_t}.




\bibliographystyle{apj}
\bibliography{Papers}

\end{document}